\documentclass{aa}
\usepackage{graphicx}
\usepackage{bm}
\usepackage{times}
\usepackage{amsmath}
\usepackage{amssymb}
\usepackage{natbib}
\usepackage{color}
\usepackage{mathrsfs}
\usepackage[colorlinks,allcolors=blue]{hyperref}
\bibpunct{(}{)}{;}{a}{}{,}
\graphicspath{{./fig/}{./png/}}
\newcommand{\uuu}{{\bm u}}

\newcommand{\FFF}{{\bm F}}

\newcommand{\SSt}{\bm{\mathsf{S}}}
\newcommand{\SStij}{\mathsf{S}_{ij}}
\newcommand{\Eq}[1]{Eq.~(\ref{#1})}
\newcommand{\Equ}[1]{Equation~(\ref{#1})}

\newcommand{\EQ}{\begin{equation}}
\newcommand{\EN}{\end{equation}}
\newcommand{\EQA}{\begin{eqnarray}}
\newcommand{\ENA}{\end{eqnarray}}
\newcommand{\brac}[1]{\langle #1 \rangle}
\newcommand{\pd}{\partial}

\newcommand{\mean}[1]{\overline{#1}}

\newcommand{\cP}{c_{\rm P}}
\newcommand{\cV}{c_{\rm V}}
\newcommand{\cs}{c_{\rm s}}

\newcommand{\urms}{u_{\rm rms}}

\newcommand{\Ma}{{\rm Ma}}

\newcommand{\chiSGSo}{\chi_{\rm SGS}^{(1)}}

\newcommand{\Hp}{H_{\rm p}}
\newcommand{\Nu}{{\rm Nu}}
\newcommand{\Pe}{{\rm Pe}}
\newcommand{\Pra}{{\rm Pr}}

\newcommand{\PraSGSo}{{\rm Pr}_{\rm SGS}^{(1)}}

\newcommand{\Ra}{{\rm Ra}}

\newcommand{\Rey}{{\rm Re}}

{}

\newcommand{\calR}{{\cal R}}
\newcommand{\Fenth}{\mean{F}_{\rm enth}}
\newcommand{\vFenth}{\mean{\bm F}_{\rm enth}}

\newcommand{\Fconv}{F_{\rm conv}}

\newcommand{\Fbot}{F_{\rm bot}}
\newcommand{\Ftot}{F_{\rm tot}}

\newcommand{\Fn}{\mathscr{F}_{\rm n}}
\newcommand{\mFenth}{\mean{F}_{\rm enth}}
\newcommand{\mFconv}{\mean{F}_{\rm conv}}
\newcommand{\mFrad}{\mean{F}_{\rm rad}}
\newcommand{\mFkin}{\mean{F}_{\rm kin}}
\newcommand{\mFvisc}{\mean{F}_{\rm visc}}

\newcommand{\mFcool}{\mean{F}_{\rm cool}}
\newcommand{\nabad}{\nabla_{\rm ad}}
\newcommand{\nabrad}{\nabla_{\rm rad}}
\newcommand{\namnad}{\nabla-\nabla_{\rm ad}}

\newcommand{\tbot}{{\rm bot}}

\newcommand{\zbot}{z_{\rm bot}}
\newcommand{\zcz}{z_{\rm CZ}}
\def\onethird{{\textstyle{1\over3}}}

\def\onehalf{{\textstyle{1\over2}}}

\newcommand{\Figas}[2]{Figs.~\ref{#1} and \ref{#2}}
\newcommand{\Fig}[1]{Figure~\ref{#1}} 
\newcommand{\Figu}[1]{Figure~\ref{#1}}

\newcommand{\Sec}[1]{Section~\ref{#1}} 
\newcommand{\Table}[1]{Table~\ref{#1}}

\begin{document}

\authorrunning{K\"apyl\"a}
\titlerunning{Overshooting in simulations of compressible convection}

   \title{Overshooting in simulations of compressible convection}

   \author{P. J. K\"apyl\"a
          \inst{1,2}
          }

   \institute{Georg-August-Universit\"at G\"ottingen, Institut f\"ur Astrophysik,
              Friedrich-Hund-Platz 1, D-37077 G\"ottingen, Germany
              \email{pkaepyl@uni-goettingen.de}
         \and ReSoLVE Centre of Excellence, Department of Computer Science,
              Aalto University, PO Box 15400, FI-00076 Aalto, Finland
}

\date{\today}

\abstract{
     Convective motions that overshoot into regions that are formally convectively stable
     cause extended mixing.
   }%
   {%
     We aim to determine the scaling of the overshooting depth ($d_{\rm os}$) at the base
     of the convection zone as a function of imposed energy flux ($\mathscr{F}_{\rm n}$)
     and to estimate the extent of overshooting at the base of the solar convection
     zone.
   }%
   {%
     Three-dimensional Cartesian simulations of hydrodynamic compressible non-rotating
     convection with unstable and stable layers were used. The
     simulations used either a fixed heat conduction profile or a
     temperature- and density-dependent formulation based on Kramers opacity law.
     The simulations covered a range of almost four orders of magnitude in the imposed
     flux, and the sub-grid scale diffusivities were varied so as to
     maintain approximately constant supercriticality at each
     flux.
   }%
   {%
     A smooth heat conduction
     profile (either fixed or through Kramers opacity law) leads to a
     relatively shallow power law with
     $d_{\rm os}\propto \mathscr{F}_{\rm n}^{0.08}$ for low $\Fn$. A fixed step-profile of the heat
     conductivity at
     the bottom of the convection zone leads to a somewhat steeper dependency
     on $d_{\rm os}\propto \mathscr{F}_{\rm n}^{0.12}$ in the same regime.
     Experiments with and without subgrid-scale entropy diffusion
       revealed a strong dependence on the effective Prandtl number, which is
       likely to explain the steep power laws as a function of $\Fn$ reported
       in the literature.
     Furthermore, changing the heat conductivity artificially in the
     radiative and overshoot layers to speed up thermal saturation is
     shown to lead to a substantial underestimation of the overshooting
     depth.
   }%
   {%
     Extrapolating from the results obtained with smooth heat conductivity
     profiles, which are the most realistic set-up we considered, suggest
     that the overshooting depth for the solar energy flux is
       about 20\%\ of
     the pressure scale height at the base of the convection zone. This
      is two to four times higher than the estimates from
     helioseismology. However, the current simulations do not include
     rotation or magnetic fields, which are known to reduce convective overshooting.
   }%


  \maketitle


\section{Introduction}

Convective mixing in stars has important consequences, for example, in
early and late phases of stellar evolution and for the diffusion of
light elements. Furthermore, stellar differential rotation
\citep[e.g.][]{R89} and dynamos \citep[e.g.][]{M78,KR80,BS05} owe
their existence to turbulent fluid motions. The efficiency of mixing
in convective and radiative layers in stars differs greatly. The
former are vigorously mixed on a timescale much shorter than the
evolutionary timescale of the star. Thus it is of great interest to be
able to predict where effective convective mixing occurs. The greatest
uncertainty in this respect is the amount of overshooting from
convection zones (CZ) to adjacent radiative layers.

Stellar structure and evolution models most often apply some variant
of the mixing length (ML) model of \cite{Vi53} to describe
convection. These models are completely local and do not allow
  overshooting. Non-local extensions to ML models
\citep[e.g.][]{1973ApJ...184..191S,1984ApJ...282..316S,1991A&A...241..227S}
yield estimates of overshooting, but the validity of the ML approach has
been questioned \citep[e.g.][]{1987A&A...188...49R}. More advanced closures of convection based on Reynolds stress
\citep[e.g.][]{1985A&A...150..133X,2006ApJ...643..426D,GOMS10,Ca11a}
are physically more consistent but challenging to implement
\citep[see,
  however,][]{2012ApJ...759L..14Z,2013ApJS..205...18Z}. Furthermore,
testing and validation of the Reynolds stress models, for example, by
comparison to three-dimensional numerical simulations, is still in its infancy
\citep[e.g.][]{1999ApJ...526L..45K,2015AN....336...32S,2018ApJ...868...12C}.

A seemingly attractive option to study overshooting is to solve the
governing equations directly by means of three-dimensional
simulations. Numerical simulations of convection have been used to
estimate the overshooting depth in numerous studies
\citep[e.g.][]{HTM86,1993A&A...277...93R,1994ApJ...421..245H,1995A&A...295..703S,1998A&A...340..178S,2000ApJ...529..402S,BCT02,2003A&A...401..433Z,2006ApJ...653..765R,2009MNRAS.398.1011T,2017A&A...604A.125P,2017ApJ...836..192B,2017ApJ...843...52H,2019MNRAS.484.1220K}. Early
studies indicated overshooting of the order of a pressure scale height
at the base of the convection zone, which is an order of magnitude more
than typical estimates from helioseismology
\citep[e.g.][]{1997MNRAS.288..572B}. The difference between these
studies and the Sun is that the energy flux imposed in the simulations
is typically much greater than the corresponding solar flux. This
leads to higher convective velocities in the simulations and to
an overestimation of the overshooting. Scaling laws, based on the relation of
convective velocities with the energy flux, arise in analytic models
of overshooting \citep[e.g.][]{Za91,2004ApJ...607.1046R} and predict a
reduction of the overshooting depth as a function of decreasing flux.
The primary aim of the current study is to establish these
relations from carefully controlled numerical experiments.

Compressible simulations with realistic solar energy flux are hampered
by the disparity of the acoustic and dynamical timescales, or by a low
Mach number, in the deep parts of the convection zone. According to
mixing length arguments, the enthalpy flux $\vFenth = \cP \mean{(\rho
  \uuu)' T'}$, where $\cP$ is the heat capacity at constant pressure,
$\rho$ is the density, $\uuu$ is the convective velocity, and $T$ is
the temperature, and the apostrophes (overbar) denote fluctuations
(averages) that can be approximated as $\Fenth=\phi \mean{\rho} u'^3$
\citep{Br16}. Assuming that $\Fenth \approx F_\odot$, it is possible
to construct a normalised energy flux \citep[e.g.][]{BCNS05},
\begin{equation}
\Fn^{(\odot)} = F_\odot/\rho \cs^3 \approx \phi \rho u'^3/\rho \cs^3 \approx \Ma^3.
\label{equ:FnMa}
\end{equation}
The last approximation is justified by the fact that the factor $\phi$
has been reported to be in the range $4-20$ \citep{BCNS05,Br16},
whereas convection carries only a fraction of the total flux in the
lower part of the solar convection zone \citep[see, e.g. the
  solar model of][]{Stix02}. Using solar values at $r_0=0.71~R_\odot$,
where $R_\odot=7\cdot10^8$~m is the solar radius,
$F_\odot=L_\odot/(4\pi r_0^2)\approx 6\cdot10^{25}$~W~m$^{-2}$, where
$L_\odot=3.84\cdot 10^{26}$~W is the solar luminosity,
$\rho\approx200$~kg~m$^{-3}$, and $\cs\approx200$~km~s$^{-1}$, the
normalised flux at the base of the solar CZ is $\Fn^{(\odot)}\approx
4\cdot 10^{-11}$. Thus the Mach number in the deep convection zone is
about $10^{-4}$. This leads to a short time-step due to the
high sound speed and to prohibitively long integration times
\citep[e.g.][]{2017LRCA....3....1K}. Although anelastic methods
\citep[e.g][]{1969JAtS...26..448G} bypass the acoustic time-step
problem at solar luminosity, any simulation attempting to do this
self-consistently would need to be run for a Kelvin--Helmholtz time to
achieve thermal saturation. This is not feasible for solar parameters
without resorting, for example, to arbitrarily changing the heat
conductivity in the radiative layer
\citep[e.g.][]{2017ApJ...836..192B} with current and any foreseeable
supercomputers \citep[e.g.][]{2017LRCA....3....1K}.

Recently, \cite{2017ApJ...843...52H} presented results from numerical
simulations of fully compressible convection where the overshooting
depth was computed from a range of two orders of magnitude in the
input flux. He reached values of $\Fn=5\cdot10^{-7}$, which is at the
limits of current numerical feasibility, and obtained a power law
$d_{\rm os}/\Hp \propto \Fn^{0.31}$ for the overshooting depth $d_{\rm
  os}$. This led \cite{2017ApJ...843...52H} to estimate that the
overshooting at the base of the solar convection zone is about 0.4\%\ of the pressure scale height, or roughly
200~km. Earlier numerical studies
\citep[e.g.][]{1998A&A...340..178S,2009MNRAS.398.1011T} have reported results
that suggest a similar steep dependency of $d_{\rm os}$ on $\Fn$.

Here these studies are revisited by a set-up where the heat
conductivity is self-consistently computed using the Kramers opacity
law. This set-up allows the depth of the convection zone to dynamically
adapt to changes in the thermodynamic state of the system
\citep{2019GApFD.113..149K} and to produce a smooth transition between
convective and radiative layers
\citep{2000gac..conf...85B,2017ApJ...845L..23K}. Furthermore, a
significantly broader range of imposed flux values is covered
than in any of the previous studies. Moreover, particular care is taken
to isolate the effect of the input flux by performing models where the
supercriticality of convection, degree of turbulence, and effective
thermal Prandlt number are approximately constant as the flux
varies. An effort is made to link to the earlier studies of
\cite{1998A&A...340..178S} and \cite{2009MNRAS.398.1011T} by targeted
sets of simulations probing the influence of subgrid scale entropy
diffusion on convection and the resulting overshooting depth.
Finally, a critical assessment of some of the modeling choices of
\cite{2017ApJ...843...52H} is presented.

The {\sc Pencil Code}\footnote{\url{https://github.com/pencil-code/}}
was used to produce the simulations. At the core of the code is a
switchable finite-difference solver for partial differential equations
that can be used to study a wide selection of physical problems. In
the current study a third-order Runge-Kutta time-stepping method and
centred sixth-order finite differences for spatial derivatives are
used \citep[cf.][]{B03}.

\section{The model} \label{sect:model}

The Kramers set-ups used in the current study are similar to those in
\cite{2017ApJ...845L..23K}. The equations for compressible
hydrodynamics
\begin{eqnarray}
\frac{D \ln \rho}{D t} &=& -\bm\nabla \bm\cdot \uuu, \label{equ:dens}\\
\frac{D\uuu}{D t} &=& {\bm g} -\frac{1}{\rho}(\bm\nabla p - \bm\nabla \bm\cdot 2 \nu \rho \bm{\mathsf{S}}),\label{equ:mom} \\
T \frac{D s}{D t} &=& -\frac{1}{\rho} \left[\bm\nabla \bm\cdot \left(\FFF_{\rm rad} + \FFF_{\rm SGS}\right) \right] + 2 \nu \bm{\mathsf{S}}^2 + \Gamma_{\rm cool},
\label{equ:ent}
\end{eqnarray}
are solved, where $D/Dt = \pd/\pd t + \uuu\cdot\bm\nabla$ is
the advective
derivative, $\rho$ is the density, $\uuu$ is the velocity,
$\bm{g}=-g\hat{\bm{e}}_z$ is the acceleration due to gravity with
$g>0$, $p$ is the pressure, $T$ is the temperature, $s$ is the
specific entropy, and $\nu$ is the constant kinematic viscosity.
Furthermore, $\FFF_{\rm rad}$ and $\FFF_{\rm SGS}$ are the radiative
and turbulent subgrid scale (SGS) fluxes, respectively, and
$\Gamma_{\rm cool}$ describes cooling at the surface (see
below). $\SSt$ is the traceless rate-of-strain tensor with
\begin{eqnarray}
\SStij = \onehalf (u_{i,j} + u_{j,i}) - \onethird \delta_{ij} \bm\nabla\cdot\uuu.
\end{eqnarray}
An optically thick fully ionised gas is considered, where radiation
is modelled through diffusion approximation. The ideal gas equation of
state $p= (\cP - \cV) \rho T =\calR \rho T$ applies, where $\calR$ is
the gas constant, and $c_{\rm V}$ is the specific heat at constant
volume. The radiative
flux is given by
\begin{eqnarray}
\FFF_{\rm rad} = -K\bm\nabla T,
\label{equ:Frad}
\end{eqnarray}
where $K$ is the radiative heat conductivity. Two qualitatively
different heat conductivity prescriptions are considered, where $K$
either has a fixed profile $K(z)$ or is a function of density and
temperature, $K(\rho,T)$. In the latter case, $K$ is given by
\begin{eqnarray}
K = \frac{16 \sigma_{\rm SB} T^3}{3 \kappa \rho},
\label{equ:Krad1}
\end{eqnarray}
where $\sigma_{\rm SB}$ is the Stefan-Boltzmann constant and $\kappa$
is the opacity. $\kappa$ is assumed to obey a power law
\begin{eqnarray}
\kappa = \kappa_0 (\rho/\rho_0)^a (T/T_0)^b,
\label{equ:kappa}
\end{eqnarray}
where $\rho_0$ and $T_0$ are reference values of density and
temperature. Equations~(\ref{equ:Krad1}) and (\ref{equ:kappa})
combine into
\begin{eqnarray}
K(\rho,T) = K_0 (\rho/\rho_0)^{-(a+1)} (T/T_0)^{3-b}.
\label{equ:Krad2}
\end{eqnarray}
Here $a=1$ and $b=-7/2$ are used, which correspond to the Kramers
opacity law for free-free and bound-free transitions
\citep{WHTR04}. Heat conductivity consistent with the Kramers law was
first used in convection simulations by \cite{2000gac..conf...85B}.

Owing to the strong depth dependence of the radiative diffusivity,
$\chi=K/(\cP\rho)$, additional turbulent SGS
diffusivity is used in the entropy equation to keep the
simulations numerically feasible. Here the SGS flux is formulated as
\begin{eqnarray}
\FFF_{\rm SGS} = -\rho T \chiSGSo \bm\nabla s',
\label{equ:FSGS}
\end{eqnarray}
where $s'=s-\mean{s}$ is the fluctuation of the specific entropy. The
overbar indicates horizontal averaging here and in what follows. The
coefficient$\chiSGSo$ is constant in the whole domain. $\FFF_{\rm
  SGS}$ has a negligible contribution to the net horizontally
averaged energy flux, such that $\mean{\FFF}_{\rm SGS} \approx
0$. The current SGS formulation is similar to those used in \cite{KKST07}, \cite{BBBMT10}, and
\cite{2019GApFD.113..149K}, for example.

The cooling at the surface is described by
\begin{eqnarray}
\Gamma_{\rm cool} = - \Gamma_0 f(z) (T_{\rm cool} - T),
\label{equ:cool}
\end{eqnarray}
where $\Gamma_0$ is a cooling luminosity, $T=e/c_{\rm V}$ is the
temperature where $e$ is the internal energy, and where $T_{\rm
  cool}=T_{\rm top}$ is a reference temperature corresponding to the
fixed value at the top boundary.

\subsection{Geometry, and initial and boundary conditions}

The computational domain is a rectangular box where $z_{\rm bot} \leq
z \leq z_{\rm top}$ is the vertical coordinate, where $z_{\rm
  bot}/d=-0.45$, $z_{\rm top}/d=1.05$, and where $d$ is the depth of
the initially isentropic layer (see below). In a few runs the domain
extends to deeper layers such that $z_{\rm bot}/d = - 0.75$ to accommodate
deeper overshooting. The horizontal coordinates $x$ and $y$ run from
$-2d$ to $2d$. The horizontal size of the box is thus $L_{\rm H}/d=4$,
and the vertical extent $L_z/d$ is either $1.5$ or $1.8$.

The initial stratification consists of three layers. Two
  configurations of the three-layer set-up are considered here: in the
  first set-up (hereafter P$^2$I), the two lower layers are polytropic
with polytropic indices $n_1=3.25$ ($z_{\rm bot}/d \leq z/d \leq 0$)
and $n_2=1.5$ ($0 \leq z/d \leq 1$). The uppermost layer above $z/d=1$
is initially
isothermal. This layer mimics a photosphere where radiative
  cooling is efficient. The choice of $n_1$ is motivated by fact that
in the
special case where the temperature gradient in the corresponding
hydrostatic state is constant, the solution is a polytrope with index
$13/4$; see \cite{BB14} and Appendix~A of \cite{Br16}. Assuming that
an extended stable layer forms at the bottom of the domain, its
stratification is close to the hydrostatic solution \citep[see,
  e.g.][]{2019GApFD.113..149K}. This, however, can only be confirmed a
posteriori because the depth of the convective layer is not pre-determined
in the cases where the Kramers opacity law is used. In the
  second set-up (hereafter P$^3$) all three layers are polytropic with
  indices $(n_1,n_2,n_3)=(2,1.5,1.5)$. In these cases the radiative
  diffusion in the uppermost layer is enhanced and no explicit cooling
  is applied. This configuration is the same as in
  \cite{1998A&A...340..178S} and was chosen to accommodate 
  comparisons with that study. The choice of $n_2=1.5$ for the thermal
  stratification in the middle layer comes from the expectation that
  the convectively unstable layer is nearly isentropic in the final
  statistically saturated state. The initial
velocity follows a Gaussian-noise distribution with an amplitude of about $10^{-4}\sqrt{dg}$.

The horizontal boundaries are periodic, and on the vertical boundaries,
impenetrable and stress-free boundary conditions are imposed for the
flow such that
\begin{eqnarray}
\frac{\pd u_x}{\pd z} = \frac{\pd u_y}{\pd z} = u_z = 0.
\end{eqnarray}
The temperature gradient at the bottom boundary is set according to
\begin{eqnarray}
\frac{\pd T}{\pd z} = -\frac{F_\tbot}{K_\tbot},
\end{eqnarray}
where $F_{\rm bot}$ is the fixed input flux and $K_\tbot(x,y,\zbot)$ is the
value of the heat conductivity at the bottom of the domain. On the
upper boundary a constant temperature $T=T_{\rm top}$, coinciding with
the initial value, is assumed.

\subsection{Units, control parameters, and simulation strategy}

The units of length, time, density, and entropy are given by
\begin{eqnarray}
[x] = d,\ \ \ [t] = \sqrt{d/g},\ \ \ [\rho] = \rho_0,\ \ \ [s] = \cP,
\end{eqnarray}
where $\rho_0$ is the initial value of the density at $z=z_{\rm top}$. The
models with Kramers heat conductivity are fully defined by
choosing the value of the kinematic
viscosity $\nu$, the gravitational acceleration $g$, the values of
$a$, $b$, $K_0$, $\rho_0$, $T_0$ and the SGS Prandtl number
\begin{eqnarray}
\PraSGSo = \frac{\nu}{\chiSGSo},
\end{eqnarray}
along with $z$-dependent cooling profile $f(z)$. The values of $K_0$,
$\rho_0$, and $T_0$ are subsumed into a new variable $\widetilde{K}_0=K_0
\rho_0^{a+1} T_0^{b-3}$ that is fixed by assuming the radiative flux
at $\zbot$ to equal $\Fbot$ in the initial state. The profile $f(z)=1$ above
$z/d=1$ and $f(z)=0$ below $z/d=1$, connecting smoothly over the
interface over a width of $0.025d$. Furthermore, $\xi_0=\Hp^{\rm
  top}/d = \mathcal{R}T_{\rm top}/gd$ sets the initial pressure scale
height at the surface, thus determining the initial density
stratification. All of the current simulations have
$\xi_0=0.054$.

In runs where a fixed profile of heat conductivity is used, the
profile $K(z)$ is needed instead of specifying $\widetilde{K}_0$.
In these cases the value of $K$ at $z=z_{\rm bot}$
is fixed similarly as was done for $\widetilde{K}_0$ in the Kramers
cases. In these cases the initial profile of the Prandtl number
  based on the radiative heat conductivity
\begin{eqnarray}
  \Pr(z) = \frac{\nu}{\chi(z)},
\end{eqnarray}
where $\chi(z)=K(z)/\cP \rho(z)$, sets the relative importance of
viscous to temperature diffusion. We note that in general
$\Pr=\Pr({\bm x},t)$ because $\rho=\rho({\bm x},t)$. In cases where
$K$ has a piecewise constant profile, it can be represented in terms
of polytropic indices $(n'_1,n'_2,n'_3),$ which refer to a
corresponding non-convective hydrostatic solution. Starting the
simulations from such solutions for the thermodynamic quantities is,
however, impractical especially if the value $n_2$ is far away from
the final convective state, which is always close to the adiabatic
value of 1.5. Thus these indices typically differ from those used to
initialise the thermal variables \citep[see also][]{BCNS05}. The heat
conductivities in the different layers are connected through
\begin{eqnarray}
\frac{K_i}{K_j} = \frac{n'_i+1}{n'_j+1},
\end{eqnarray}
where $i$ and $j$ refer to any of the three layers.

The dimensionless normalised flux is given by
\begin{eqnarray}
  \mathscr{F}_{\rm n} = \Fbot/\rho_\tbot c_{\rm s,bot}^3,
\end{eqnarray}
where $\rho_\tbot$ and $c_{\rm s,bot}$ are the density and the sound
speed, respectively, at $z_{\rm bot}$ in the initial non-convecting
state.

The input energy flux determines the overall convective velocity
realised in the simulations via $u_{\rm conv} \propto \mathscr{F}_{\rm
  n}^{1/3}$, see \Eq{equ:FnMa} and \cite{2018arXiv180709309K}. Thus if
only $\mathscr{F}_{\rm n}$ were changed, the relative importance
of the diffusion coefficients, measured by Reynolds and P\'eclet
numbers, would change as well. To eliminate these dependences, and to be able
to concentrate solely on the effects of varying input flux, the
viscosity and SGS entropy diffusion are scaled proportional to
$\mathscr{F}_{\rm n}^{1/3}$. In addition to changing the diffusion
coefficients, the cooling luminosity $\Gamma_0$ at the surface is also
scaled proportionally to $\mathscr{F}_{\rm n}$. The input flux itself
is varied by changing the overall magnitude of $K$ such that the value
at the bottom is given by $K_{\rm bot}=-\Fbot/(\pd T/\pd z)_{z_{\rm
    bot}}$. Furthermore, the degree of supercriticality of convection,
measured by a Rayleigh number, and the Prandtl number, describing the
ratio of viscosity to thermal diffusion, will affect the properties of
convection and overshooting if they are allowed to vary. The current
choice of the dependency of $\nu$ and $\chiSGSo$ on $\Fn$, however,
eliminates these dependences such that the supercriticality and
effective Prandtl numbers are nearly constant over the range of $\Fn$
considered here. This is discussed in detail in \Sec{subsec:basic}.

In addition to the explicit viscosity, SGS, and radiative diffusion,
the advective terms in each of the \Equ{equ:dens}--(\ref{equ:ent}) are
written in terms of a fifth-order upwinding derivative with a
hyperdiffusive sixth-order correction where the diffusion coefficient
depends locally on the flow, see Appendix~B of \cite{DSB06}.

\subsection{Diagnostics quantities}

The following quantities are outcomes of the simulations that can only
be determined a posteriori. These include the global Reynolds and
SGS P\'eclet numbers,
\begin{eqnarray}
\Rey = \frac{\urms}{\nu k_1},\ \ \ \Pe_{\rm SGS} = \frac{\urms}{\chi_{\rm SGS}^{(1)} k_1},
\end{eqnarray}
where $\urms$ is the volume-averaged rms velocity, and $k_1=2\pi/d$ is
an estimate of the largest eddies in the system. Typically
$\chi\ll\chiSGSo$ in the CZ in most of the current
simulations. Furthermore, $\chi$ has a strong depth dependence due to
the density stratification. Thus it is useful to define Reynolds and effective P\'eclet numbers separately for the overshoot zone,
\begin{eqnarray}
\Rey_{\rm OZ} = \frac{\urms^{\rm OZ}}{\nu k_{\rm OZ}},\ \ \Pe^{\rm eff}_{\rm OZ} = \frac{\urms^{\rm OZ}}{(\chi_{\rm SGS}^{(1)}+\mean{\chi}_{\rm OZ}) k_{\rm OZ}},
\end{eqnarray}
where all quantities are taken from the base of the
convection zone, $z=z_{\rm CZ}$, with $\mean{\chi}_{\rm
  OZ}=\mean{K}_{\rm OZ}/(c_{\rm P}\mean{\rho}_{\rm OZ})$ being the
mean (horizontally averaged) radiative diffusivity, and where
$k_{\rm OZ}=2\pi/d_{\rm os}$ is a wavenumber based on the depth of the
overshoot layer $d_{\rm os}$.
Similarly, an effective Prandtl number can be defined as
\begin{eqnarray}
\Pra^{\rm eff}_{\rm OZ} = \frac{\nu}{(\chi_{\rm SGS}^{(1)}+\mean{\chi}_{\rm OZ})}.
\end{eqnarray}
Precise definitions of $\zcz$ and $d_{\rm os}$ are given in
\Sec{sect:definitions}.

To assess the level of supercriticality of convection, radiative and
SGS Rayleigh numbers are defined as
\begin{eqnarray}
\Ra^{\rm Rad} &=& \frac{gd^4}{\nu \chi}\left( - \frac{1}{\cP}\frac{{\rm d}s}{{\rm d}z} \right),\\
\Ra^{\rm SGS} &=& \frac{gd^4}{\nu \chi_{\rm SGS}^{(1)}}\left( - \frac{1}{\cP}\frac{{\rm d}s}{{\rm d}z} \right).
\end{eqnarray}
Supercriticality of convection is roughly determined by
$\mbox{min}(\Ra^{\rm Rad},\Ra^{\rm SGS})$. Both
quantities vary as functions of height and are quoted near the surface
at $z/d=0.85$ for all models. Conventionally, the Rayleigh number in
the hydrostatic non-convecting state is one of the control
parameters. This is still true for the cases with fixed $K$ profile in
the current study, but in the runs with Kramers conductivity, the
convectively unstable layer in the hydrostatic case is very thin and
confined to the near-surface layers \citep{Br16}. Thus the Rayleigh
numbers are quoted from the thermally saturated and statistically
stationary states.

The path in parameter space taken here is artificial in that in real
fluids, there is no SGS diffusivity ($\Ra^{\rm SGS}=0$) and any change
in the flux would be reflected by the then decisive radiative Rayleigh
number. For example, a system where $\Pra$ is fixed $\Ra^{\rm
  rad}\propto\Fn^{-4/3}$. However, taking this path severely limits
the computationally feasible range of $\Fn$ because the Reynolds and
P\'eclet number also increase in proportion to $\Fn$ and the
resolution requirements quickly become prohibitive.

\begin{table*}[t!]
\centering
\caption[]{Summary of the runs with smoothly varying profiles of $K$.}
  \label{tab:runs1}
       \vspace{-0.5cm}
      $$
          \begin{array}{p{0.05\linewidth}ccccccccccc}
          \hline
          \hline
          \noalign{\smallskip}
          Run & \Fn & \Ra^{\rm Rad} [10^7] & \Ra^{\rm SGS} [10^6] & \Pr_{\rm OZ}^{\rm eff} & \Rey & \Rey_{\rm OZ} & \Pe_{\rm OZ}^{\rm eff} & \Nu & \Delta\rho & d_{\rm os}/\Hp & K \\
          \hline
          K-3 &  4.5\cdot 10^{-4} & 0.12 &  0.5  &  0.4  &  15  &  1.3  &  0.5  &  19  &   55  &  1.29  & \mbox{Kramers} \\ 
          K-2 &  1.8\cdot 10^{-4} & 0.64 &  0.6  &  0.5  &  17  &  1.2  &  0.7  &  36  &   74  &  1.05  & \mbox{Kramers} \\ 
          K-1 &  9.1\cdot 10^{-5} &  1.7 &  0.8  &  0.7  &  18  &  1.2  &  0.8  &  51  &   91  &  0.90  & \mbox{Kramers} \\ 
          K0  &  4.5\cdot 10^{-5} &  4.0 &  0.9  &  0.8  &  19  &  1.1  &  0.9  &  63  &  108  &  0.79  & \mbox{Kramers} \\ 
          K1  &  1.8\cdot 10^{-5} &  9.9 &  1.0  &  0.9  &  20  &  1.0  &  0.8  &  77  &  128  &  0.69  & \mbox{Kramers} \\ 
          K2  &  9.1\cdot 10^{-6} &   18 &  1.1  &  0.9  &  20  &  0.9  &  0.8  &  84  &  139  &  0.63  & \mbox{Kramers} \\ 
          K3  &  4.6\cdot 10^{-6} &   28 &  1.1  &  0.9  &  20  &  0.9  &  0.8  &  88  &  147  &  0.60  & \mbox{Kramers} \\ 
          K4  &  1.8\cdot 10^{-6} &   55 &  1.1  &  1.0  &  19  &  0.8  &  0.8  &  92  &  154  &  0.55  & \mbox{Kramers} \\ 
          K5  &  9.1\cdot 10^{-7} &   94 &  1.1  &  1.0  &  19  &  0.8  &  0.8  &  94  &  158  &  0.51  & \mbox{Kramers} \\ 
          K6  &  4.6\cdot 10^{-7} &  157 &  1.2  &  1.0  &  19  &  0.8  &  0.7  &  96  &  161  &  0.49  & \mbox{Kramers} \\ 
          K7  &  1.8\cdot 10^{-7} &  302 &  1.2  &  1.0  &  19  &  0.8  &  0.8  &  98  &  164  &  0.46  & \mbox{Kramers} \\ 
          \hline
          K-3h & 4.6\cdot 10^{-4} & 0.39 &  2.4  &  0.2  &  34  &  2.8  &  0.7  &  24  &   58  &  1.28  & \mbox{Kramers} \\ 
          K-2h & 1.8\cdot 10^{-4} &  1.7 &  3.1  &  0.4  &  37  &  2.8  &  1.1  &  41  &   76  &  1.04  & \mbox{Kramers} \\ 
          K-1h & 9.1\cdot 10^{-5} &  4.3 &  3.6  &  0.5  &  39  &  2.7  &  1.4  &  55  &   93  &  0.88  & \mbox{Kramers} \\ 
          K0h  & 4.5\cdot 10^{-5} &  9.6 &  4.2  &  0.6  &  41  &  2.5  &  1.6  &  68  &  111  &  0.77  & \mbox{Kramers} \\ 
          K1h  & 1.8\cdot 10^{-5} &   23 &  4.7  &  0.7  &  42  &  2.3  &  1.7  &  80  &  130  &  0.67  & \mbox{Kramers} \\ 
          K2h  & 9.1\cdot 10^{-6} &   40 &  4.8  &  0.8  &  42  &  2.1  &  1.7  &  87  &  141  &  0.61  & \mbox{Kramers} \\ 
          K3h  & 4.6\cdot 10^{-6} &   65 &  4.7  &  0.9  &  41  &  2.0  &  1.8  &  91  &  150  &  0.58  & \mbox{Kramers} \\ 
          K4h  & 1.8\cdot 10^{-6} &  130 &  5.0  &  0.9  &  41  &  2.0  &  1.9  &  94  &  157  &  0.53  & \mbox{Kramers} \\ 
          K5h  & 9.1\cdot 10^{-7} &  221 &  5.3  &  1.0  &  40  &  2.0  &  1.9  &  97  &  161  &  0.50  & \mbox{Kramers} \\ 
          K6h  & 4.6\cdot 10^{-7} &  367 &  5.5  &  1.0  &  40  &  1.9  &  1.8  &  98  &  163  &  0.47  & \mbox{Kramers} \\ 
          K7h  & 1.8\cdot 10^{-7} &  700 &  5.6  &  1.0  &  40  &  1.7  &  1.7  &  99  &  166  &  0.45  & \mbox{Kramers} \\ 
          \hline
          P-1 &  8.9\cdot 10^{-5} & 0.27 &  0.8  &  0.7  &  19  &  1.3  &  0.9  &  71  &   86  &  0.86  & \mbox{K-profile} \\ 
          P0  &  4.5\cdot 10^{-5} & 0.47 &  0.9  &  0.8  &  19  &  1.1  &  0.9  &  71  &  106  &  0.76  & \mbox{K-profile} \\ 
          P1  &  1.8\cdot 10^{-5} & 0.92 &  1.0  &  0.9  &  20  &  1.0  &  0.9  &  71  &  127  &  0.68  & \mbox{K-profile} \\ 
          P2  &  9.1\cdot 10^{-6} &   15 &  1.1  &  0.9  &  20  &  0.9  &  0.8  &  71  &  139  &  0.63  & \mbox{K-profile} \\ 
          P3  &  4.5\cdot 10^{-6} &   23 &  1.1  &  0.9  &  19  &  0.9  &  0.8  &  71  &  146  &  0.61  & \mbox{K-profile} \\ 
          P4  &  1.8\cdot 10^{-6} &   42 &  1.1  &  1.0  &  19  &  0.8  &  0.8  &  71  &  154  &  0.56  & \mbox{K-profile} \\ 
          P5  &  9.1\cdot 10^{-7} &   71 &  1.2  &  1.0  &  19  &  0.8  &  0.8  &  71  &  158  &  0.52  & \mbox{K-profile} \\ 
          P6  &  4.6\cdot 10^{-7} &  116 &  1.2  &  1.0  &  19  &  0.8  &  0.8  &  71  &  161  &  0.49  & \mbox{K-profile} \\ 
          \hline
          \end{array}
          $$
          \tablefoot{
            The Nusselt number is quoted from $z/d=0.85$ and
              $\Pr_{\rm OZ}^{\rm eff}$ refers to the effective Prandtl
              number at $z/d=0$ here and in
              all following tables. The grid resolution in all runs is $288^3$.}
\end{table*}

\begin{figure*}
  \includegraphics[width=\textwidth]{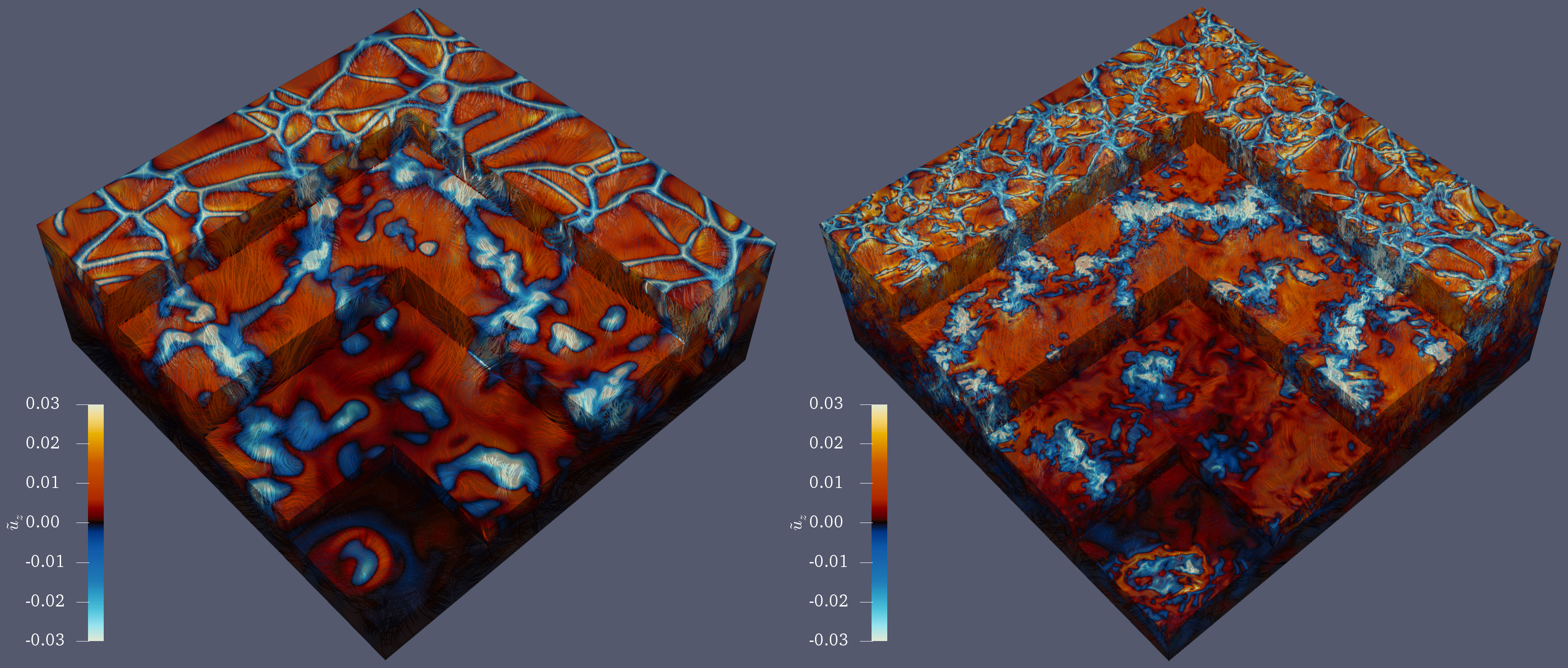}
\caption{Normalised vertical velocity $\tilde{u}_z=u_z/(gd)^{1/2}$
  (colours) and streamlines of the flow from Runs~R4 (left) and R7
  (right).}
\label{fig:uuz_boxes}
\end{figure*}

\begin{table*}[t!]
\centering
\caption[]{Summary of the runs with a fixed step profile of $K$.}
  \label{tab:runs2}
       \vspace{-0.5cm}
      $$
          \begin{array}{p{0.05\linewidth}ccccccccccc}
          \hline
          \hline
          \noalign{\smallskip}
          Run & \Fn & \Ra^{\rm Rad} [10^6] & \Ra^{\rm SGS} [10^6] & \Pr_{\rm OZ}^{\rm eff} & \Rey & \Rey_{\rm OZ} & \Pe_{\rm OZ}^{\rm eff} & \Nu & \Delta\rho & d_{\rm os}/\Hp & K \\
          \hline
          S-1 & 5.0\cdot 10^{-5} &  1.5  &   1.3  &  0.7  &  17  &   1.3  &   1.0  &   2.5  &  150  &  0.71  & \mbox{Step} \\ 
          S0  & 2.5\cdot 10^{-5} &  1.9  &   1.1  &  0.8  &  17  &   1.1  &   0.9  &   2.5  &  174  &  0.61  & \mbox{Step} \\ 
          S1  & 1.8\cdot 10^{-5} &  2.9  &   1.0  &  0.9  &  19  &   1.0  &   0.9  &   2.5  &  135  &  0.54  & \mbox{Step} \\ 
          S2  & 8.9\cdot 10^{-6} &  4.2  &   0.9  &  0.9  &  19  &   0.9  &   0.8  &   2.5  &  145  &  0.50  & \mbox{Step} \\ 
          S3  & 4.5\cdot 10^{-6} &  6.2  &   0.9  &  0.9  &  18  &   0.9  &   0.8  &   2.5  &  150  &  0.47  & \mbox{Step} \\ 
          S4  & 1.8\cdot 10^{-6} &   11  &   0.9  &  1.0  &  18  &   0.8  &   0.8  &   2.5  &  156  &  0.42  & \mbox{Step} \\ 
          S5  & 9.0\cdot 10^{-7} &   18  &   0.9  &  1.0  &  17  &   0.8  &   0.7  &   2.5  &  161  &  0.39  & \mbox{Step} \\ 
          S6  & 4.5\cdot 10^{-7} &   29  &   0.9  &  1.0  &  17  &   0.7  &   0.7  &   2.5  &  165  &  0.35  & \mbox{Step} \\ 
          S7  & 1.8\cdot 10^{-7} &   52  &   0.9  &  1.0  &  17  &   0.6  &   0.6  &   2.5  &  169  &  0.33  & \mbox{Step} \\ 
          \hline
          S-1h & 5.0\cdot 10^{-5} &  2.8 &   4.9  &  0.6  &  37  &   3.0  &   1.8  &   2.5  &  151  &  0.71  & \mbox{Step} \\ 
          S0h  & 2.5\cdot 10^{-5} &  3.8 &   4.5  &  0.7  &  37  &   2.6  &   1.8  &   2.5  &  175  &  0.62  & \mbox{Step} \\ 
          S1h  & 1.8\cdot 10^{-5} &  6.1 &   4.2  &  0.8  &  42  &   2.3  &   1.8  &   2.5  &  138  &  0.52  & \mbox{Step} \\ 
          S2h  & 8.9\cdot 10^{-6} &  9.0 &   4.0  &  0.9  &  42  &   2.2  &   1.9  &   2.5  &  146  &  0.48  & \mbox{Step} \\ 
          S3h  & 4.5\cdot 10^{-6} &   13 &   3.7  &  0.9  &  41  &   2.0  &   1.8  &   2.5  &  153  &  0.44  & \mbox{Step} \\ 
          S4h  & 1.8\cdot 10^{-6} &   24 &   3.8  &  0.9  &  39  &   1.9  &   1.8  &   2.5  &  159  &  0.39  & \mbox{Step} \\ 
          S5h  & 9.0\cdot 10^{-7} &   39 &   3.9  &  1.0  &  38  &   1.7  &   1.7  &   2.5  &  162  &  0.37  & \mbox{Step} \\ 
          S6h  & 4.5\cdot 10^{-7} &   63 &   3.9  &  1.0  &  38  &   1.5  &   1.5  &   2.5  &  166  &  0.33  & \mbox{Step} \\ 
          \hline
          \end{array}
          $$
          \tablefoot{The grid resolution was $288^3$ for all other runs except for
            S-1, S0, S-1h, and S0h, which have a deeper domain
            extending to $z/d=-0.75$ and a $336^3$ grid, and Run~S7, which
            was run with a $144^3$ grid. All runs have $(n'_1,n'_2,n'_3)=(3.25,0,0)$.}
\end{table*}

\begin{table*}[t!]
\centering
\caption[]{Summary of the runs with a double-step profile for $K$.}
  \label{tab:runsDS}
       \vspace{-0.5cm}
      $$
          \begin{array}{p{0.05\linewidth}ccccccccccc}
          \hline
          \hline
          \noalign{\smallskip}
          Run & \Fn & \Ra^{\rm Rad} [10^6] & \Ra^{\rm SGS} [10^6] & \Pr_{\rm OZ}^{\rm eff} & \Rey & \Rey_{\rm OZ} & \Pe_{\rm OZ}^{\rm eff} & \Nu & \Delta\rho & d_{\rm os}/\Hp & K \\
          \hline
          DS0   & 1.0\cdot 10^{-4} &  4.8  &  -  &  3.7  & 17  &  1.6  & 6.2  &  25  &  23  &  0.68  & \mbox{2--Step} \\ 
          DS1   & 4.1\cdot 10^{-5} &   10  &  -  &  6.7  & 16  &  0.8  & 5.8  &  25  &  29  &  0.52  & \mbox{2--Step} \\ 
          DS2   & 2.1\cdot 10^{-5} &   18  &  -  &   10  & 16  &  0.6  & 6.6  &  25  &  33  &  0.44  & \mbox{2--Step} \\ 
          DS3   & 1.0\cdot 10^{-5} &   30  &  -  &   16  & 16  &  0.5  & 8.1  &  25  &  36  &  0.37  & \mbox{2--Step} \\ 
          DS4   & 4.2\cdot 10^{-6} &   57  &  -  &   27  & 16  &  0.4  &  10  &  25  &  39  &  0.28  & \mbox{2--Step} \\ 
          DS5   & 2.1\cdot 10^{-6} &   91  &  -  &   41  & 16  &  0.3  &  12  &  25  &  41  &  0.23  & \mbox{2--Step} \\ 
          DS5h  & 2.1\cdot 10^{-6} &   92  &  -  &   41  & 16  &  0.2  & 9.4  &  25  &  41  &  0.22  & \mbox{2--Step} \\ 
          \hline
          DSS10 & 1.0\cdot 10^{-4} &  4.5  & 1.6 &  2.2  & 18  &  3.3  &  7.4 &  25  &  22  &  (0.76)  & \mbox{2--Step} \\ 
          DSS11 & 4.1\cdot 10^{-5} &  9.6  & 2.0 &  2.9  & 19  &  2.7  &  8.1 &  25  &  28  &  (0.78)  & \mbox{2--Step} \\ 
          DSS12 & 2.0\cdot 10^{-5} &   17  & 2.3 &  3.5  & 19  &  2.3  &  8.1 &  25  &  32  &  (0.77)  & \mbox{2--Step} \\ 
          DSS13 & 1.0\cdot 10^{-5} &   29  & 2.6 &  3.9  & 19  &  1.9  &  7.4 &  25  &  35  &  0.74  & \mbox{2--Step} \\ 
          DSS14 & 4.1\cdot 10^{-6} &   57  & 2.8 &  4.3  & 19  &  1.6  &  7.0 &  25  &  38  &  0.70  & \mbox{2--Step} \\ 
          DSS15 & 2.1\cdot 10^{-6} &   95  & 3.0 &  4.5  & 19  &  1.3  &  5.7 &  25  &  39  &  0.64  & \mbox{2--Step} \\ 
          \hline
          DSS20 & 1.0\cdot 10^{-4} &  4.5  & 3.3 &  2.9  & 18  &  2.8  &  8.1 &  25  &  22  &  (0.74)  & \mbox{2--Step} \\ 
          DSS21 & 4.1\cdot 10^{-5} &  9.5  & 4.0 &  4.3  & 18  &  2.0  &  8.3 &  25  &  28  &  (0.73)  & \mbox{2--Step} \\ 
          DSS22 & 2.0\cdot 10^{-5} &   17  & 4.6 &  5.3  & 18  &  1.5  &  7.8 &  25  &  32  &  0.67  & \mbox{2--Step} \\ 
          DSS23 & 1.0\cdot 10^{-5} &   29  & 5.1 &  6.4  & 18  &  1.2  &  7.6 &  25  &  35  &  0.61  & \mbox{2--Step} \\ 
          DSS24 & 4.1\cdot 10^{-6} &   57  & 5.6 &  7.6  & 18  &  1.0  &  7.5 &  25  &  38  &  0.54  & \mbox{2--Step} \\ 
          DSS25 & 2.1\cdot 10^{-6} &   94  & 5.9 &  8.3  & 18  &  0.9  &  5.6 &  25  &  39  &  0.50  & \mbox{2--Step} \\ 
          \hline
          \end{array}
          $$
          \tablefoot{
            The SGS diffusivity $\chi_{\rm SGS}^{(1)}=0$ in Set~DS and
            thus $\Ra^{\rm SGS}$ is not defined. In Set~DSS1
            (DSS2) $\Pr_{\rm SGS}^{(1)}=5\ (10)$. The grid
            resolution is $288^3$ in all runs except for DS5h, which
            was run with $576^3$ grid points.
            All runs have $(n'_1,n'_2,n'_3)=(2,-0.9,1.5)$. Parentheses for values of
            $d_{\rm os}/\Hp$ for Runs~DSS10--2 and DSS20--1 indicate
            that the results are affected by the lower boundary.}
\end{table*}

\begin{table*}[t!]
\centering
\caption[]{Summary of the runs with varying $\nu$ and $\chiSGSo$.}
  \label{tab:runs3}
       \vspace{-0.5cm}
      $$
          \begin{array}{p{0.05\linewidth}cccccccccc}
          \hline
          \hline
          \noalign{\smallskip}
          Run & \Ra^{\rm Rad} [10^7] & \Ra^{\rm SGS} [10^7] & \Pr_{\rm OZ}^{\rm eff} & \Rey & \Rey_{\rm OZ} & \Pe_{\rm OZ}^{\rm eff} & \Nu & \Delta\rho & d_{\rm os}/\Hp & {\rm grid} \\
          \hline
          R1  &   18 & 0.005  &  1.0  &   4  &  0.1  &  0.1  &  87  &  150  &  0.45  &  288^3  \\ 
          R2  &   44 & 0.028  &  1.0  &   9  &  0.3  &  0.3  &  92  &  155  &  0.50  &  288^3  \\ 
          R3  &   93 &  0.11  &  1.0  &  19  &  0.8  &  0.8  &  94  &  158  &  0.51  &  288^3  \\ 
          R4  &  221 &  0.53  &  1.0  &  40  &  2.0  &  1.9  &  97  &  161  &  0.50  &  288^3  \\ 
          R5  &  798 &   5.2  &  0.9  & 122  &  7.0  &  6.2  & 102  &  163  &  0.49  &  576^3  \\ 
          R6  & 1580 &    21  &  0.8  & 257  &   16  &   13  & 103  &  163  &  0.51  & 1152^3  \\ 
          R7  & 3310 &    85  &  0.7  & 523  &   37  &   34  & 105  &  166  &  0.52  & 1152^3  \\ 
          \hline
          \end{array}
          $$
          \tablefoot{
            All runs in this set use $\Fn=9.1\cdot10^{-7}$ and Kramers-based
            heat conductivity.}
\end{table*}

Contributions to the vertical energy flux are
\begin{eqnarray}
\mFrad  &=& - \mean{K} \frac{\pd \mean{T}}{\pd z},\\
\mFenth &=& \cP \mean{(\rho u_z)' T'},\\
\mFkin  &=& \onehalf \mean{\rho \uuu^2 u_z'},\\
\mFvisc &=& -2 \nu \mean{\rho u_i \mathsf{S}_{iz}}\\
\mFcool &=& \int_{z_{\rm bot}}^{z_{\rm top}} \Gamma_{\rm cool} {\rm d}z.
\end{eqnarray}
Here the primes denote fluctuations and overbars horizontal
averages. The
total convected flux \citep{CBTMH91} is the sum of the enthalpy and
kinetic energy fluxes:
\begin{eqnarray}
\mFconv = \mFenth + \mFkin.
\end{eqnarray}

The radiative flux can be written in terms of the mean
double-logarithmic temperature gradient $\nabla=\pd \ln \mean{T}/\pd
\ln \mean{p}$ as
\begin{equation}
\mFrad = -\mean{K} \frac{\pd \mean{T}}{\pd z}= \frac{\mean{K}g}{\cP}\frac{\nabla}{\nabad},
\end{equation}
where $\cP \nabad = \cP (1 - 1/\gamma) = \mathcal{R}$, and
$g=|\bm{g}|$. Above, $\nabla$ is the actual temperature gradient
realised in the system. Now it is possible to define the total flux
and the flux transported by adiabatic stratification as
\begin{equation}
\Ftot = \frac{\mean{K}g}{\cP}\frac{\nabrad}{\nabad},\ \ \ \mFrad^{\rm ad} = \frac{\mean{K}g}{\cP},
\end{equation}
where $\nabrad$ is a hypothetical radiative gradient in the absence of
convection and $\Ftot=\Fbot$. The ratio of the total-to-adiabatic flux
is the Nusselt number \citep[e.g.][]{Br16},
\begin{equation}
\Nu = \frac{\Ftot}{\mFrad^{\rm ad}} = \frac{\nabrad}{\nabad}.
\end{equation}
In the current set-up the Nusselt number is fixed from the outset for
cases with a static profile of $\mean{K}$. This is because the total
flux imposed at the lower boundary is proportional to $\mean{K}$. In
the cases where Kramers conductivity is used, $\mean{K}$ can evolve
but the change in the Nusselt number is not very large; see, for
example, \Table{tab:runs1}. The behaviour of $\Nu$ is different in
cases where fixed temperature is imposed at both boundaries
\citep[e.g.][]{2016GeoJI.204.1120Y,2016JFM...808..690G} because the
total flux is no longer in direct proportion with the heat
conductivity.

\section{Definitions of convection zone and overshooting} \label{sect:definitions}

To characterise the different layers, the nomenclature introduced in
\cite{2017ApJ...845L..23K} is used, although with somewhat differing
definitions. The CZ is defined to be the part of the
domain where $\mFconv>0$, whereas in the overshoot zone (OZ)
$\Fconv<0$. This is motivated by the work of
\cite{2008MNRAS.386.1979D}, who used a similar definition but employed
$\mFenth$ instead of $\mFconv$. A definition of the overshooting depth
based on $\mFenth$ was also used in
\cite{2017ApJ...845L..23K}. However, because the kinetic energy flux
carries a substantial fraction of the energy, it is natural to include
it in the definition of the convection zone. The bottom of the CZ is
denoted by $z_{\rm CZ}$.

The mean overshooting depth $z_{\rm OS}$ is taken to be the position
where the horizontally averaged $\mFkin$ drops below 1\%\ of
its value at $z_{\rm CZ}$. Various earlier studies have used a similar
definition
\citep{HTM86,1994ApJ...421..245H,1995A&A...295..703S,1998A&A...340..178S,BCT02}.
In most of the previous studies the location of the bottom of the CZ
is assumed to be fixed by the initial conditions. Here this assumption
is relaxed and $z_{\rm CZ}$ is computed from the simulation
data using the definition given above. The values of $z_{\rm OS}$ and
$z_{\rm CZ}$ are obtained by linear interpolation from the grid points
closest to the respective transitions. Furthermore, $z_{\rm CZ}$ and
$z_{\rm OS}$ are functions of time. The overshooting depth is
  defined as
\begin{equation}
 d_{\rm os}= \brac{z_{\rm CZ}(t)-z_{\rm OS}(t)}_t,
\end{equation}
where $\brac{\cdots}_t$ denotes a time average over the statistically
stationary part of the time series. Error estimates for $d_{\rm os}$
are obtained by dividing the time series into three equally long parts
and considering their largest deviation from the time average over the
whole time series as the error.

The radiative zone (RZ) is defined as the region below $z_{\rm OS}$
, and the buoyancy zone (BZ) is where $\mFconv>0$ and $\pd_z \mean{s} <
0$. Finally, the Deardorff zone (DZ) is characterised by a formally
stable stratification with a positive vertical mean entropy gradient
($\pd_z \mean{s}>0$) and $\mFconv>0$; see, 
\cite{2015ApJ...799..142T}. In this layer, the convective energy
transport is dominated by a non-local non-gradient contribution to the
enthalpy flux introduced by \cite{1961JAtS...18..540D,De66}; see also
\cite{Br16} and \cite{2017ApJ...845L..23K}. Such layers have been
reported by various authors from simulations
\citep[e.g.][]{CG92,1993A&A...277...93R,2015ApJ...799..142T,2017PhRvE..96c3104K,2017ApJ...851...74B,2018PhFl...30d6602K,2018ApJ...859..117N}.
Furthermore, the union of OZ, DZ, and BZ is
referred to as the mixed zone (MZ).

\section{Results} \label{sect:results}

Four main sets of simulations, denoted as K, P, S, and DS were conducted,
see Tables~\ref{tab:runs1}--\ref{tab:runsDS}. The sets are named after
their heat conductivity prescriptions: in Set~K, the Kramers law is
used to compute the heat conductivity. In Set~P, a static profile
corresponding to the heat conductivity computed according to the
Kramers law in the initial state of the simulation is used. Set~S
employs a static step profile of heat conductivity, $K=K(z)$. These
three sets correspond to Runs~K, P, and S of
\cite{2017ApJ...845L..23K}. Additionally, Sets~Kh and Sh are the
otherwise the same as Sets~K and S, but lower viscosity and SGS entropy
diffusion were used. These runs were branched off from corresponding
thermally saturated snapshots from runs in Sets~K and S,
respectively. In Sets~DS, DSS1, and DSS2 a double-step profile for
$K$, similar to that in \cite{1998A&A...340..178S}, was used. In Set~DS
$\chiSGSo=0,$  and in Set~DSS1 (DSS2) the SGS diffusion is included
with $\Pr_{\rm SGS}^{(1)}=5\ (10)$. Runs~DS5 and DS5h were the
same except for the grid resolution to study numerical convergence of
the results. Finally, in Set~R the explicit diffusivities $\nu$ and
$\chiSGSo$ were varied while all other parameters were kept fixed using
Run~K5 as a progenitor, see \Table{tab:runs3}. Sets~DS, DSS1, and DSS2
used the P$^3$ set-up, and the remaining sets were initialised with
the P$^2$I set-up.

\subsection{Basic characterisation of the solutions}
\label{subsec:basic}

Typical flow patterns for Runs~R4 and R7 are shown in
\Fig{fig:uuz_boxes}. The flow structure observed in various studies of
stratified non-rotating convection is recovered with connected
downflows near the surface merging into isolated plumes at larger depths
\citep[e.g.][]{SN89,1998ApJ...499..914S}. The downflows are surrounded
by broader upflows. In the majority of the current simulations the
flows are at best mildly turbulent with $\Rey\approx20\ldots40$ such as
in Run~R4 in the left panel
of \Fig{fig:uuz_boxes}. However, the qualitative large-scale structure
of convection does not change at higher resolutions and Rayleigh and
  Reynolds numbers, see the right panel of \Fig{fig:uuz_boxes} for Run~R7.

\begin{figure}
  \includegraphics[width=.5\textwidth]{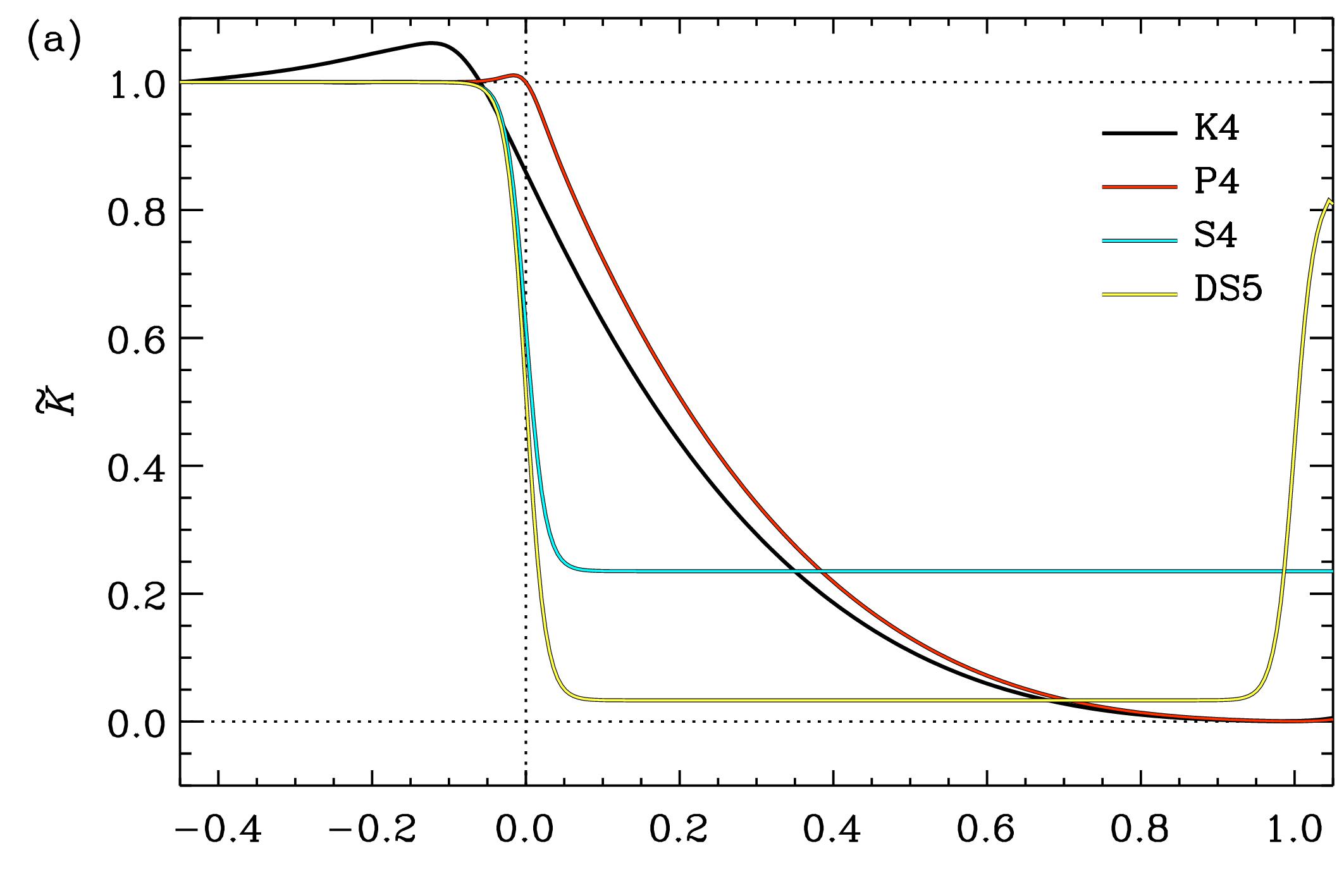}
  \includegraphics[width=.5\textwidth]{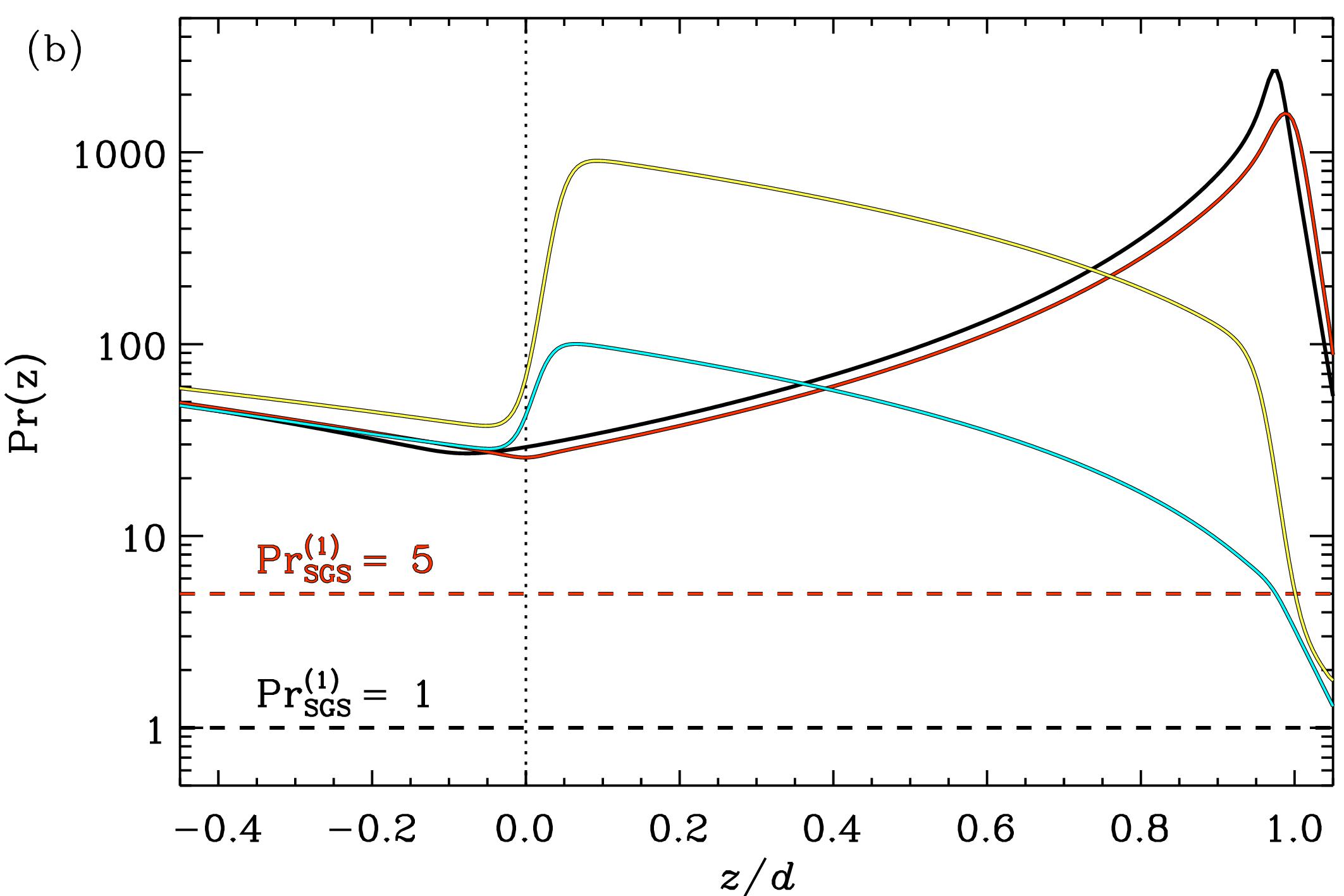}
  \caption{Horizontally averaged and normalised heat conductivity 
      $\widetilde{K}=\mean{K}(z)/K_{\rm bot}$ ({\it a}) and Prandtl
      number $\Pr(z)$ ({\it b}) from Runs~K4, P4, S4, and
      DS5. The dashed black (red) line in the lower panel
      indicates $\Pr_{\rm SGS}^{(1)}=1$ ($\Pr_{\rm SGS}^{(1)}=5$).}
\label{fig:pKappa_sets}
\end{figure}

\Figu{fig:pKappa_sets}(a) shows the profiles of $\mean{K}(z)$ from
representative runs in each of the main sets K, P, S, and DS. The
Kramers run K4 and the corresponding fixed profile model P4 have a
smoothly varying profile as a function of height with very small
values of $\mean{K}$ near the surface. In the piecewise polytropic run
S4, the profile of $\mean{K}$ is characterised by constant values in
the upper ($z>0$) and lower ($z<0$) layers
\citep[e.g.][]{HTM86,NBJRRST92}, which can be characterised by the
ratios $K_2/K_1=K_3/K_1=4/17\approx0.235$. In Run~DS5,
$K_2/K_1=2/85\approx0.0333$ and the uppermost layer above $z/d=1$ has
another constant value of $\mean{K}$ corresponding to
$K_2/K_1=5/6$. However, the transitions of $\mean{K}$ between the
layers are smoothed over a distance of $0.05d,$ due to which the value
of $\mean{K}$ corresponding to $n'_3=1.5$ is achieved only near the
upper surface.

The vertical profiles of $\Pr(z)$ and $\Pr_{\rm SGS}^{(1)}$ are shown
in \Figu{fig:pKappa_sets}(b). The strong variation of $\Pr$ as
function of depth is to be contrasted with the constant value of
$\Pr_{\rm SGS}^{(1)}$. Although $\Pr\propto\Fn$, it is still typically
greater than unity at $z/d=0$ in almost all cases, with the exception
of a few runs with the highest values of $\Fn$ (see
Tables~\ref{tab:runs1}--\ref{tab:runs3}). This is in stark contrast to
the solar convection zone, where $\Pr\ll1$ everywhere
\citep[e.g.][]{O03}.  Numerical simulations with Prandtl numbers far
different from unity are challenging numerically and render parameter
studies infeasible.  Thus an enhanced SGS diffusivity with $\Pr_{\rm
  SGS}^{(1)}$ of the order of unity was applied in most of the current
simulations. In most of the current simulations the effective Prandtl
number is dominated by the SGS diffusion.

\begin{figure}
  \includegraphics[width=.5\textwidth]{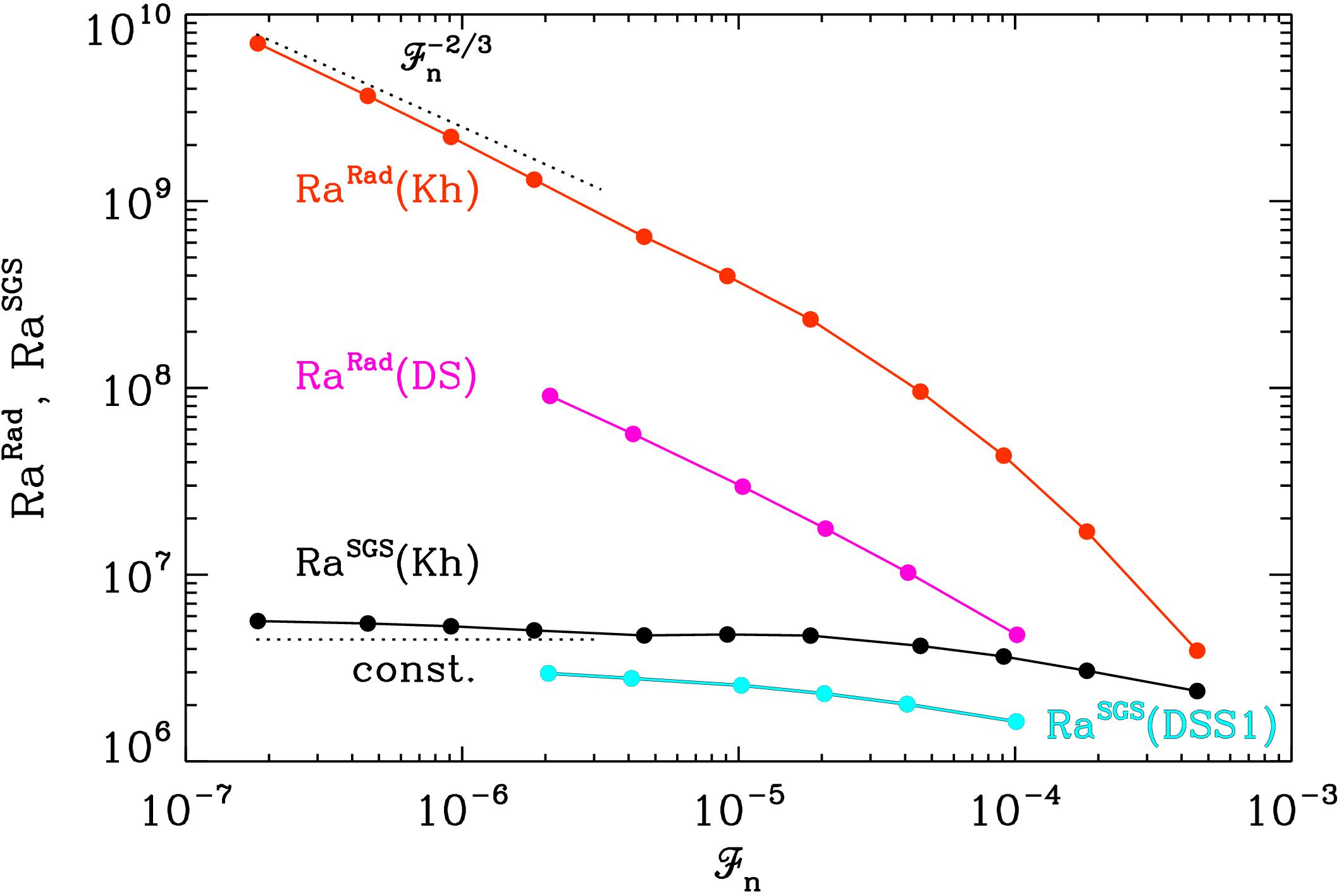}
  \caption{Rayleigh numbers $\Ra^{\rm SGS}$ (black) and
      $\Ra^{\rm Rad}$ (red) as functions of $\Fn$ from Set~Kh. The
      purple (magenta) line shows $\Ra^{\rm Rad}$ ($\Ra^{\rm SGS}$)
      from Set~DS (DSS). The dotted lines show scalings expected from
      mixing length arguments.}
\label{fig:pRayleigh}
\end{figure}

Changing the input energy flux$\Fn$ refers to varying the magnitude
of $\mean{K}$ proportionally and thus $\chi\propto\Fn$. In combination
with $\nu\propto\Fn^{1/3}$ and the mixing-length estimate $-(\Hp/\cP)
{\rm d}s/{\rm d}z = \namnad \propto \Fn^{2/3}$ \citep{Vi53}, this
involves changing $\Ra^{\rm Rad}\propto\Fn^{-2/3}$. This is to be
contrasted with the SGS Rayleigh number, where $\chi$ is replaced by
$\chi_{\rm SGS}^{(1)} \propto \Fn^{1/3}$ , which leads to $\Ra^{\rm
  SGS}$ being independent of $\Fn$. These scalings are in accordance
with the simulations, as can be
seen from \Fig{fig:pRayleigh} for Sets~Kh, DS, and DSS1. As mentioned
earlier, the supercriticality of convection is determined roughly by
${\rm min}(\Ra^{\rm Rad},\Ra^{\rm SGS})$. In the current simulations
the SGS Rayleigh number is almost always smaller. This also means that
with $\Ra^{\rm SGS}\approx$ constant, the supercriticality of
convection is also constant, and it is eliminated as an influence on
the overshooting depth. For completeness, the flux-based Rayleigh
number \citep[e.g.][]{2017LRSP...14....4B}
\begin{eqnarray}
\Ra^{\rm Flux} = \Nu\Ra^{\rm Rad} = \frac{\Ftot d^4}{\rho \chi^2 \nu}\left(-\frac{1}{\cP}\frac{{\rm d}s}{{\rm d}z}\right),
\end{eqnarray}
can be seen to vary as $\Ra^{\rm Flux}\propto\Fn^{-2/3}$ given the
dependences above. We note that $\Nu\Ra^{\rm SGS}$ is independent of
$\Fn$.

\begin{figure}
  \includegraphics[width=.5\textwidth]{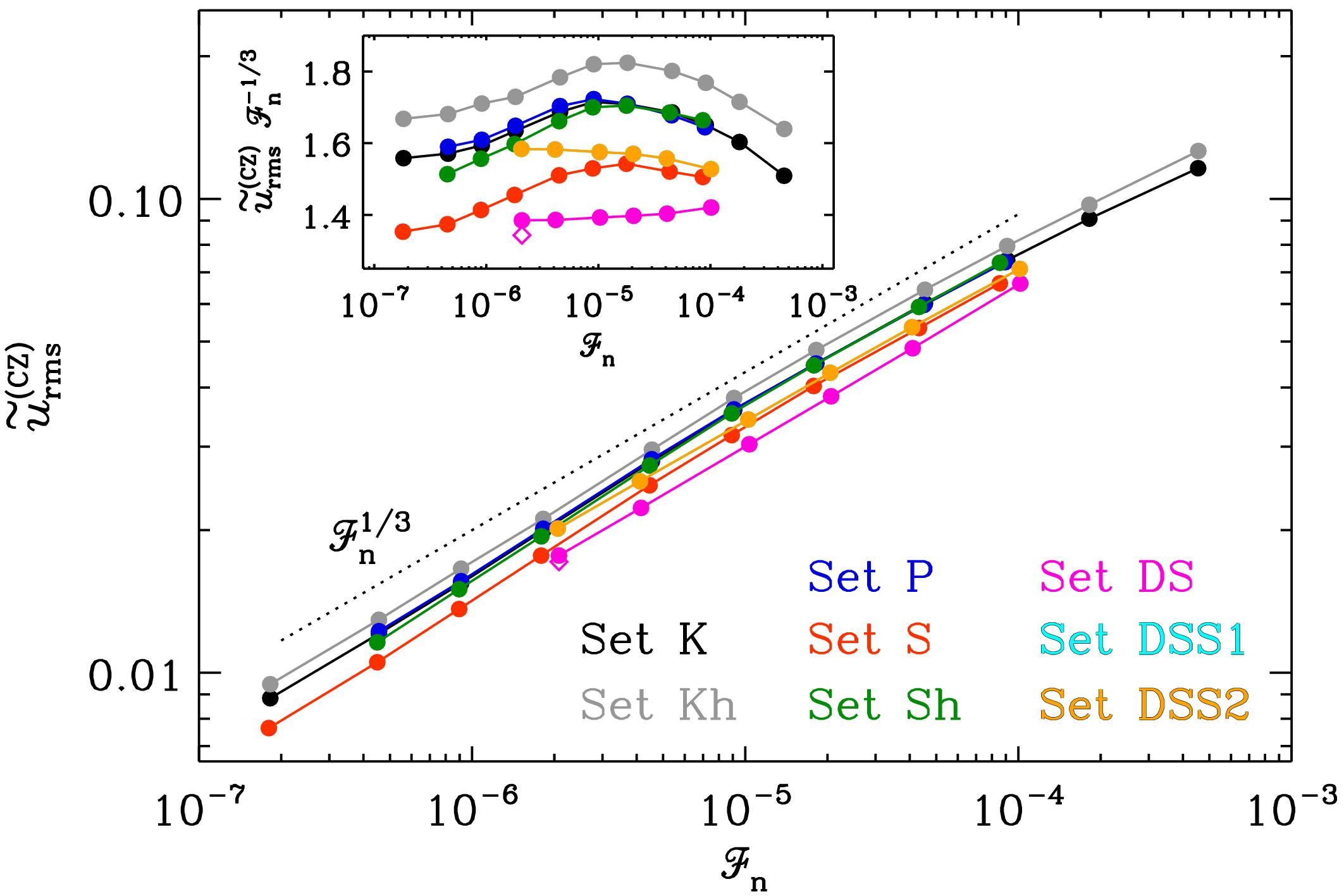}
  \caption{Normalised rms velocity $\tilde{u}_{\rm
      rms}=\urms/\sqrt{gd}$ in the CZ as a function of
    $\mathscr{F}_{\rm n}$ for the simulation sets indicated by the
    legend. The dotted line is proportional to $\mathscr{F}_{\rm
      n}^{1/3}$. The inset shows $\tilde{u}_{\rm rms} \mathscr{F}_{\rm
      n}^{-1/3}$ for the same runs. The purple diamond denotes
    Run~DS5h.}
\label{fig:purms}
\end{figure}

\Figu{fig:purms} shows the dependence of the rms velocity within the CZ from
all simulations except for those in Set~R. The simulation results are close
to a $\Fn^{1/3}$ dependence, with the coefficient of proportionality varying
between 1.3 and 1.8. This is in agreement with mixing length estimates
\citep[e.g.][]{Vi53,BCNS05,Br16} and earlier numerical findings
\citep{BCNS05,KKKBOP15,2018arXiv180709309K}.

\begin{figure*}
  \includegraphics[width=.5\textwidth]{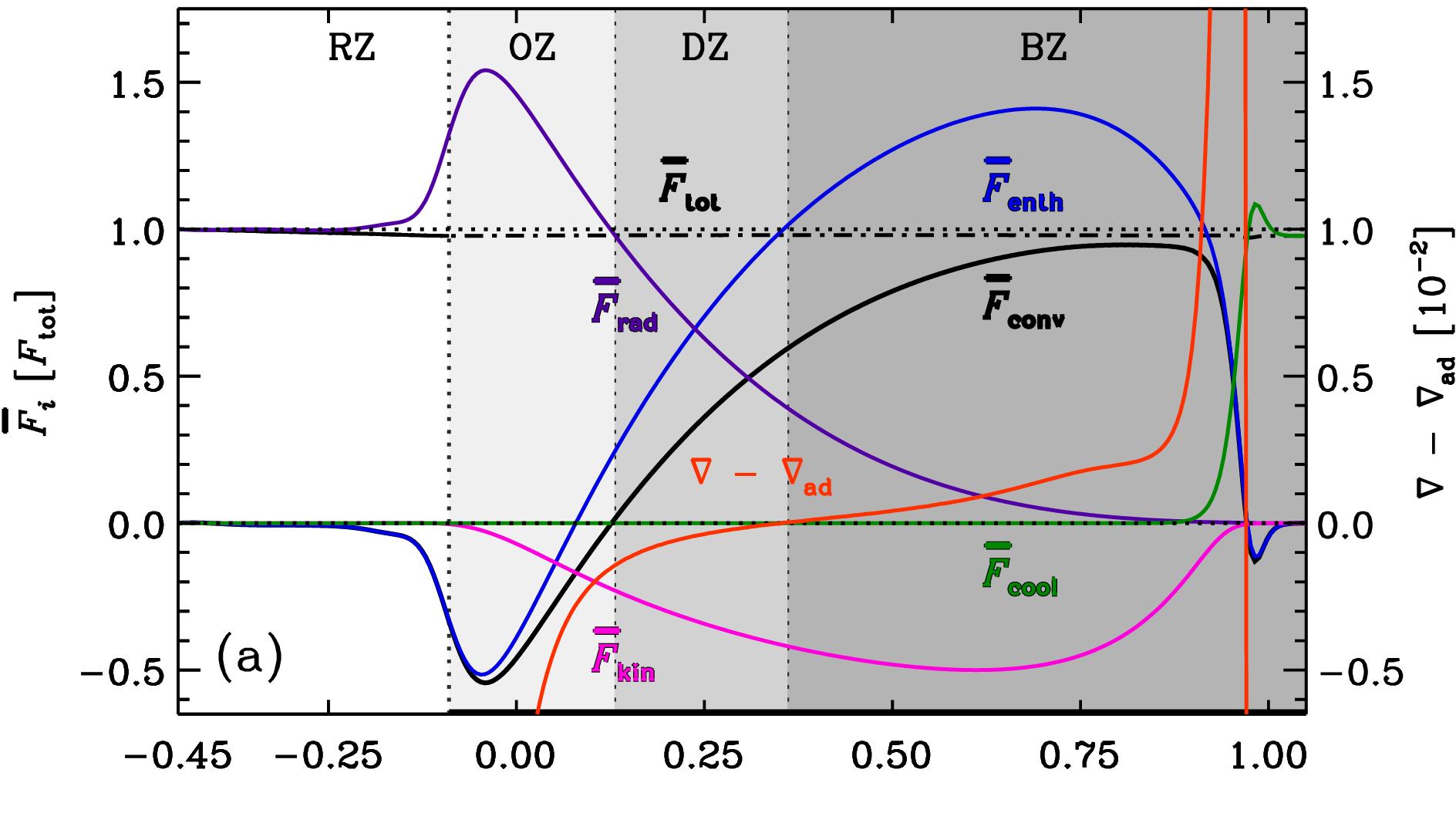}\includegraphics[width=.5\textwidth]{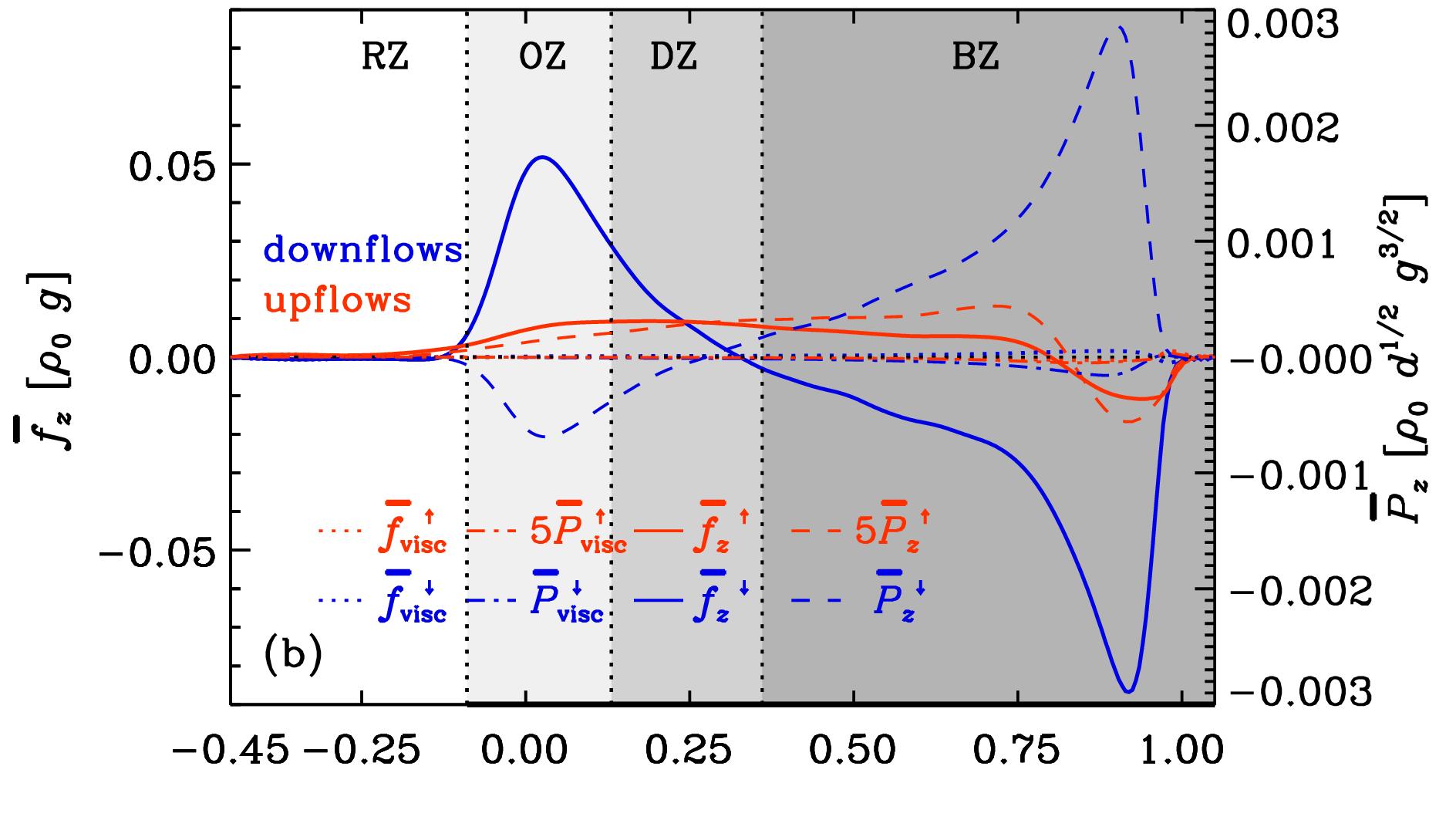}
  \includegraphics[width=.5\textwidth]{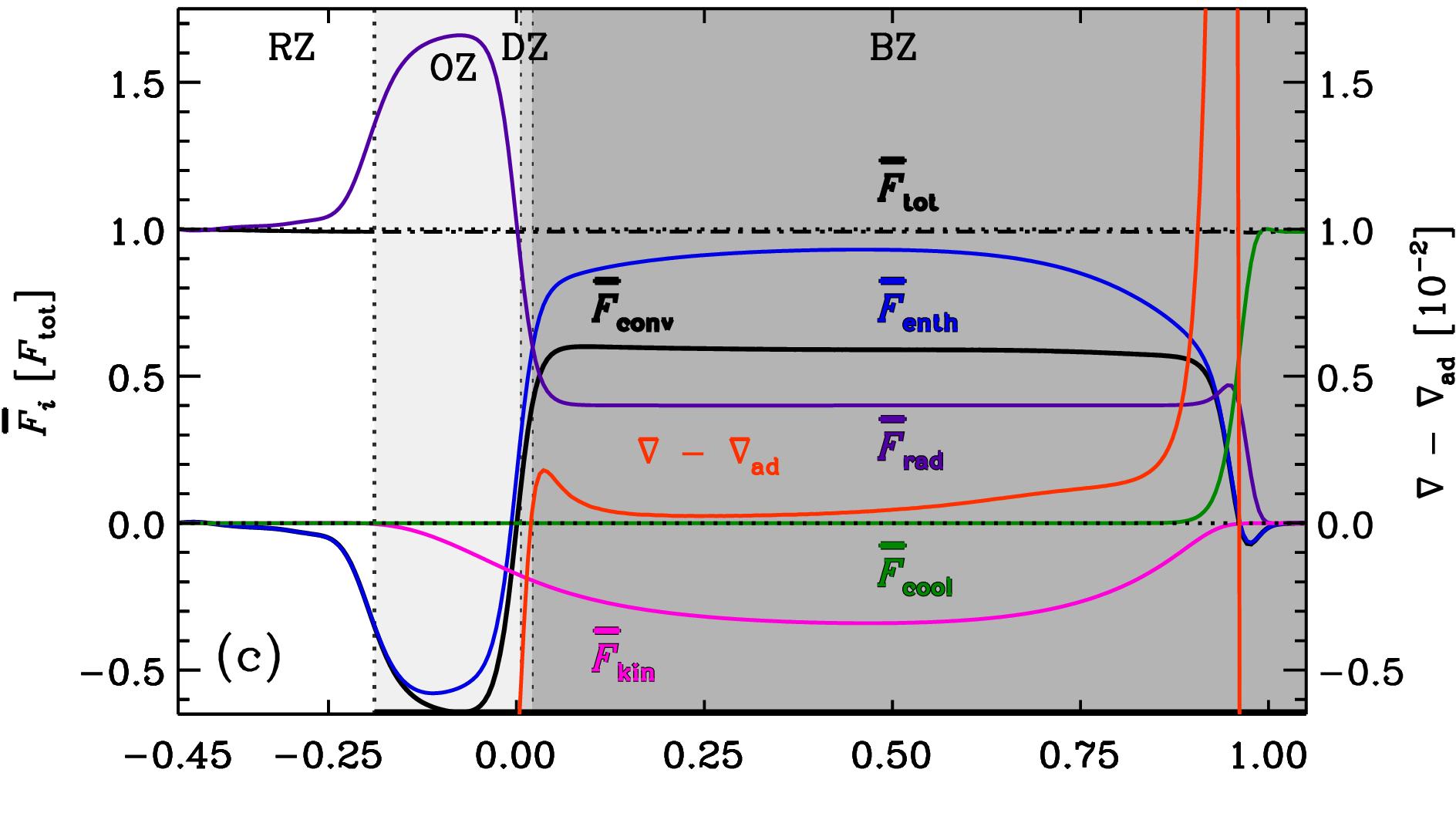}\includegraphics[width=.5\textwidth]{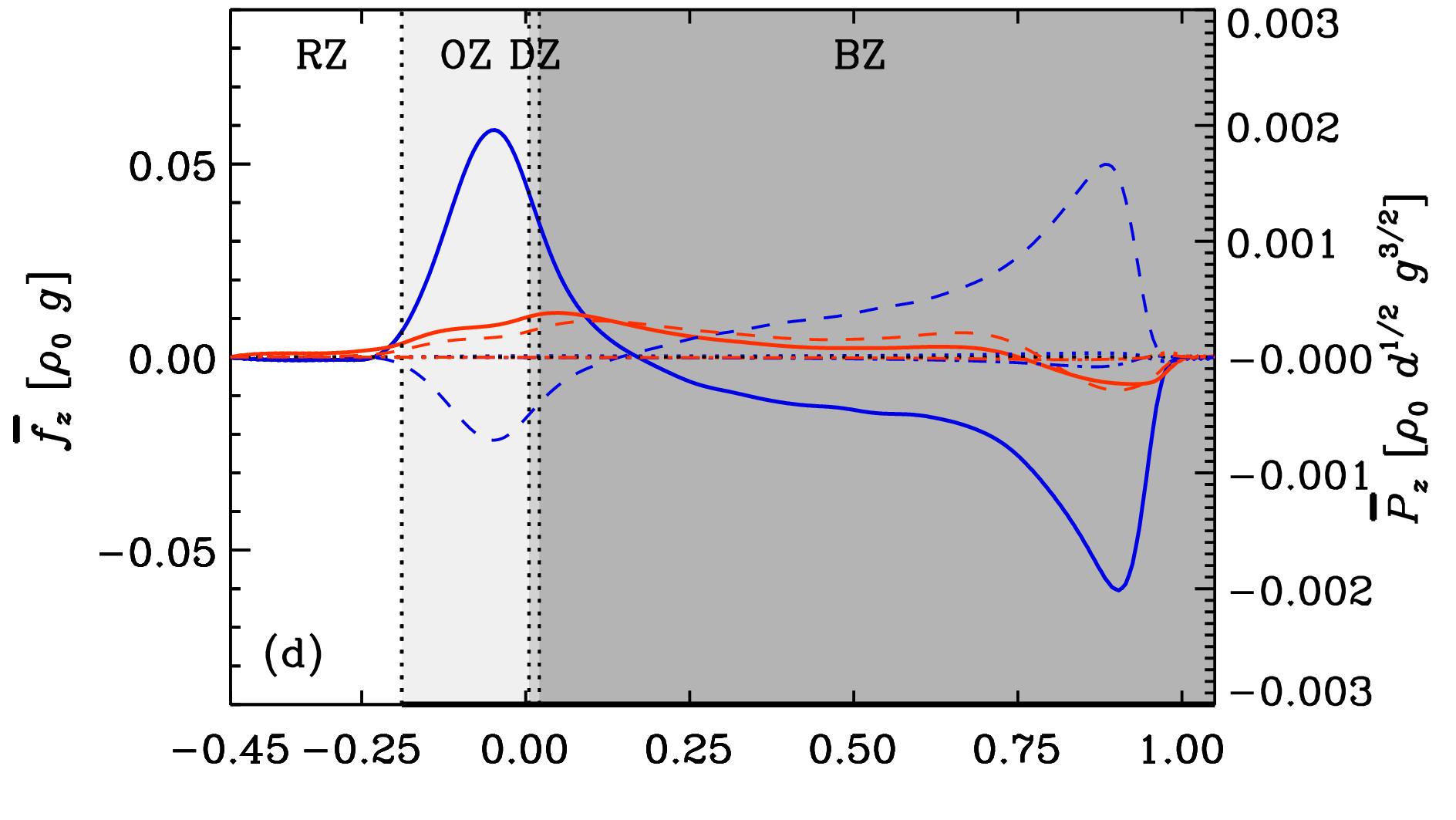}
  \includegraphics[width=.5\textwidth]{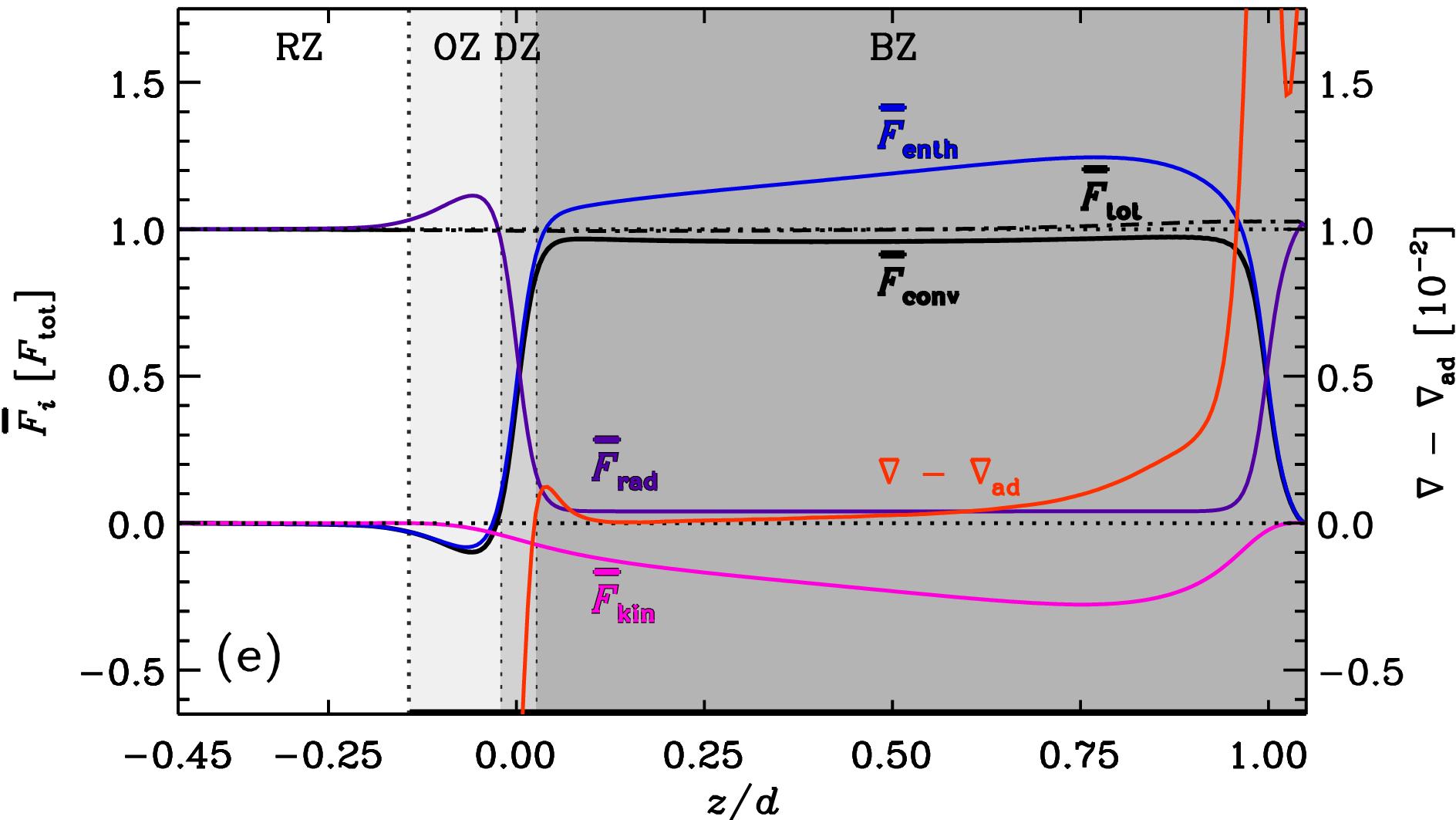}\includegraphics[width=.5\textwidth]{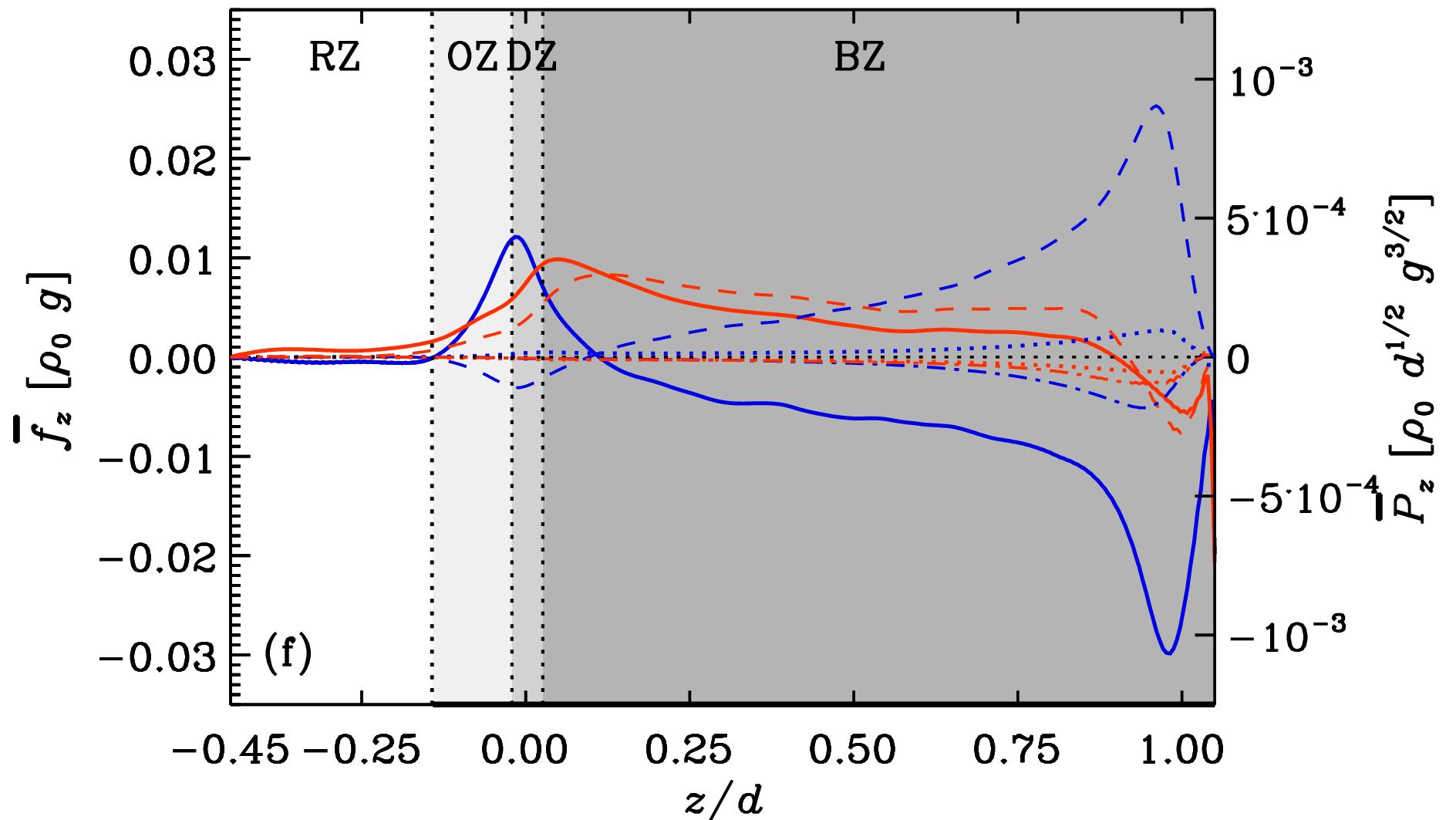}
  \caption{Panels ({\it a}), ({\it c}), and ({\it e}) show the
      total (black dash-dotted lines), convective (black),
    enthalpy (blue), kinetic energy (light purple), radiative (dark
    purple), and cooling (green) fluxes and the superadiabatic
    temperature gradient (red) from Runs~K4, S4, and DS5
,    respectively. Panels ({\it b}), ({\it d}), and ({\it f})
    show the total averaged
    vertical forces (solid lines) and the power of the forces (dashed)
    on the upflows (red) and downflows (blue) from the same runs. The
    dotted lines in these panels show the corresponding viscous force
    and its power. The shaded areas indicate the BZ (darkest), DZ, and
    OZ (lightest) and the thick black line at the horizontal axis denotes the
    MZ.}
\label{fig:flux_comp}
\end{figure*}

Results regarding the energy fluxes and force balance from
representative runs from Sets~K, S, and DS are shown in
\Fig{fig:flux_comp}. The left panels show the contributions to the
energy flux and the superadiabatic temperature gradient
$\nabla-\nabla_{\rm ad}$. The fluxes for Runs~K4 and S4 in
\Fig{fig:flux_comp}(a) and (c) are qualitatively similar to those of
Runs~K and S of \cite{2017ApJ...845L..23K}, respectively. The main
difference to the latter is the treatment of the near surface
layers; cooling layer in the present runs as opposed to an imposed
entropy gradient in \cite{2017ApJ...845L..23K}, and the somewhat
different values of $\Fn$; $1.8\cdot 10^{-5}$ here versus $9\cdot
10^{-6}$ in \cite{2017ApJ...845L..23K}. The subadiabatic Deardorff
zone encompasses roughly a quarter of the MZ in Run~K4, whereas in
Run~S4, the DZ is almost absent. The runs in Set~DS differ from those
in Set~S in that convection transports almost all of
the energy due to the lower $K_2/K_1$ ratio. Assuming that the
temperature gradient in the final statistically saturated state is
nearly adiabatic, the fraction of convective transport can be
estimated from \citep[cf.][]{BCNS05}
\begin{eqnarray}
\frac{\Fconv}{\Ftot} \approx 1 - \frac{\nabad}{\nabrad} = 1 - \Nu^{-1} = 1 - \nabad(n_2'+1).
\end{eqnarray}
According to this expression, the convective flux transports 60\%\ (96\%) in the runs in Set~S (DS). This is confirmed by the numerical
results.

Figures~\ref{fig:flux_comp}(b), (d), and (f) show the horizontally
averaged vertical total and viscous forces, $\mean{f}_z=\mean{\rho
  Du_z/Dt}$, and $\mean{f}_{\rm visc}=2\nu\pd_i(\mean{\rho S_{iz}})$,
respectively, and the resulting power of the forces
($\mean{P}_z=\mean{u_z f_z}$ and $\mean{P}_{\rm visc}=\mean{u_z f_{\rm
    visc}}$) separately for the upflows ($\uparrow$) and the downflows
($\downarrow$) from Runs~K4, S4, and DS5. The force balance in Run~K4
(\Fig{fig:flux_comp}b), is very similar to the corresponding Run~K in
\cite{2017ApJ...845L..23K}, see their Fig.~2(b). The downflows appear
to adhere to the Schwarzschild criterion such that they are
accelerated in the BZ and decelerated in the layers below. This is
contrasted by the upflows that are accelerated in the stably
stratified OZ and DZ and in the lower part of the BZ. As demonstrated
in \cite{2017ApJ...845L..23K}, the upflows are not driven by the
convective instability but are a result of matter displaced by the
deeply penetrating downflows. In Run~S4 the downflows are decelerated
already in the lower part of the BZ whereas the force on the upflows
is qualitatively similar to that in Run~K4. The force balance in
Run~DS5 is qualitatively similar to that in Run~S4 although the
magnitude and details of the quantities differ. Interestingly the sign
of the total force in the stably stratified layer near the surface is
not reversed. The shallowness of the layer is likely contributing to
this. Another aspect is the (true) overshooting from the convection
zone below. It can, however, be concluded that displacement of the
matter due to the downflows is driving the upflows in the OZ and DZ
also in Runs~S4 and DS5. However, the superadiabatic temperature
gradient has a local maximum at the bottom of the BZ in these cases,
see \Fig{fig:flux_comp}(c) and (e). Thus it is possible that the
convective instability is contributing more to the upward acceleration
in these cases. The viscous force is small in all cases and has a
noticeable effect only in the near-surface layers above $z/d=0.75$.

\subsection{Dependence of overshooting on input flux}
\label{sec:flux}

Reaching the solar value of $\Fn$ is currently not possible due to the
prohibitive time-step constraint and the long thermal adjustment time
involved
\citep[e.g.][]{2000gac..conf...85B,2017LRCA....3....1K}. However, it
is reasonable to assume that the overshooting depth scales with a
power law as a function of $\Fn$
\citep[e.g.][]{1984ApJ...282..316S,Za91}. Thus it is in principle
possible to estimate the extent of overshooting in the Sun provided
that a sufficiently broad range of higher flux values are probed and
their results are extrapolated to the solar case. A few such
studies can be found in the literature
\citep[e.g.][]{1998A&A...340..178S,2009MNRAS.398.1011T,2017ApJ...843...52H}.

\begin{figure}
  \includegraphics[width=.49\textwidth]{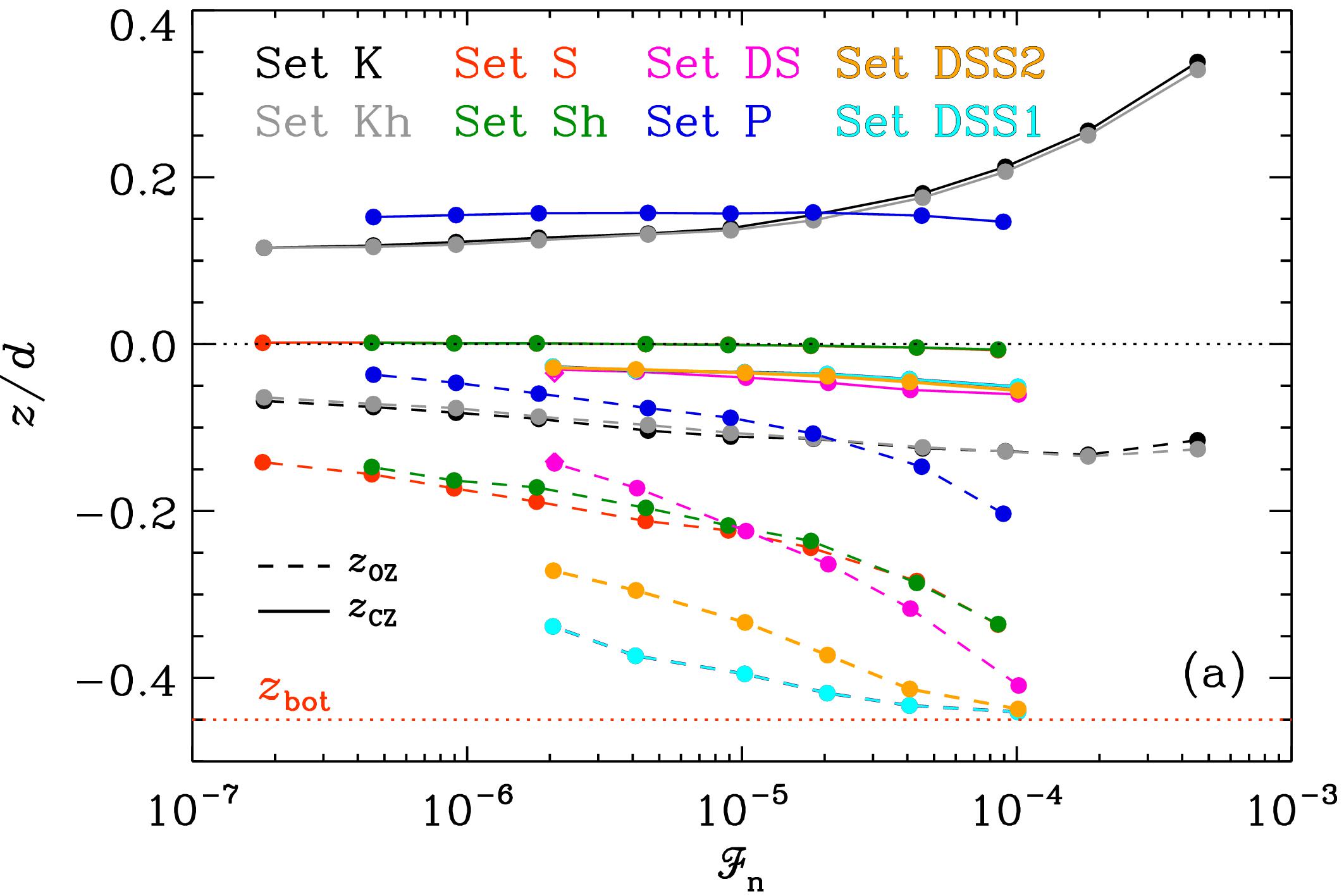}
  \includegraphics[width=.49\textwidth]{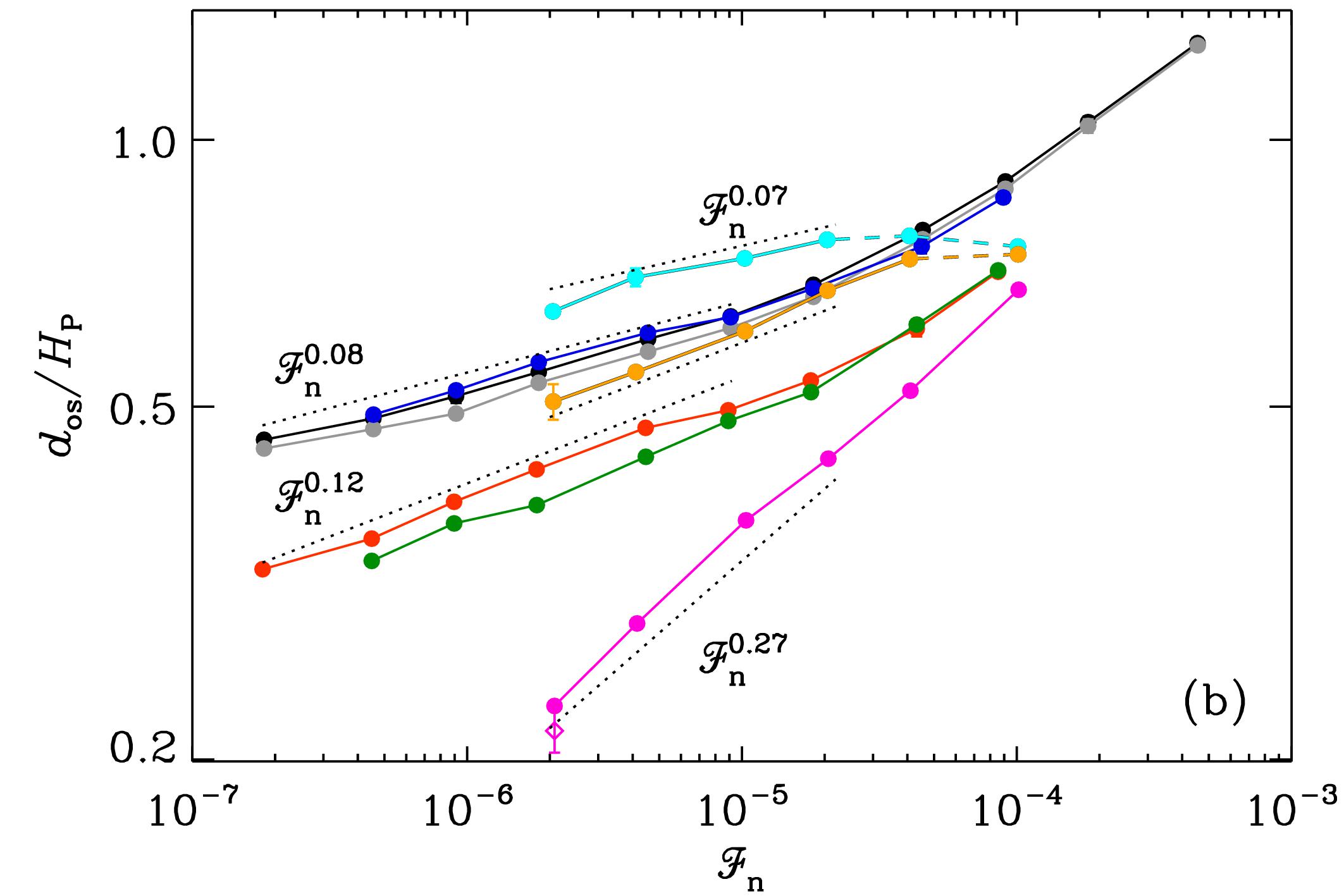}
  \includegraphics[width=.49\textwidth]{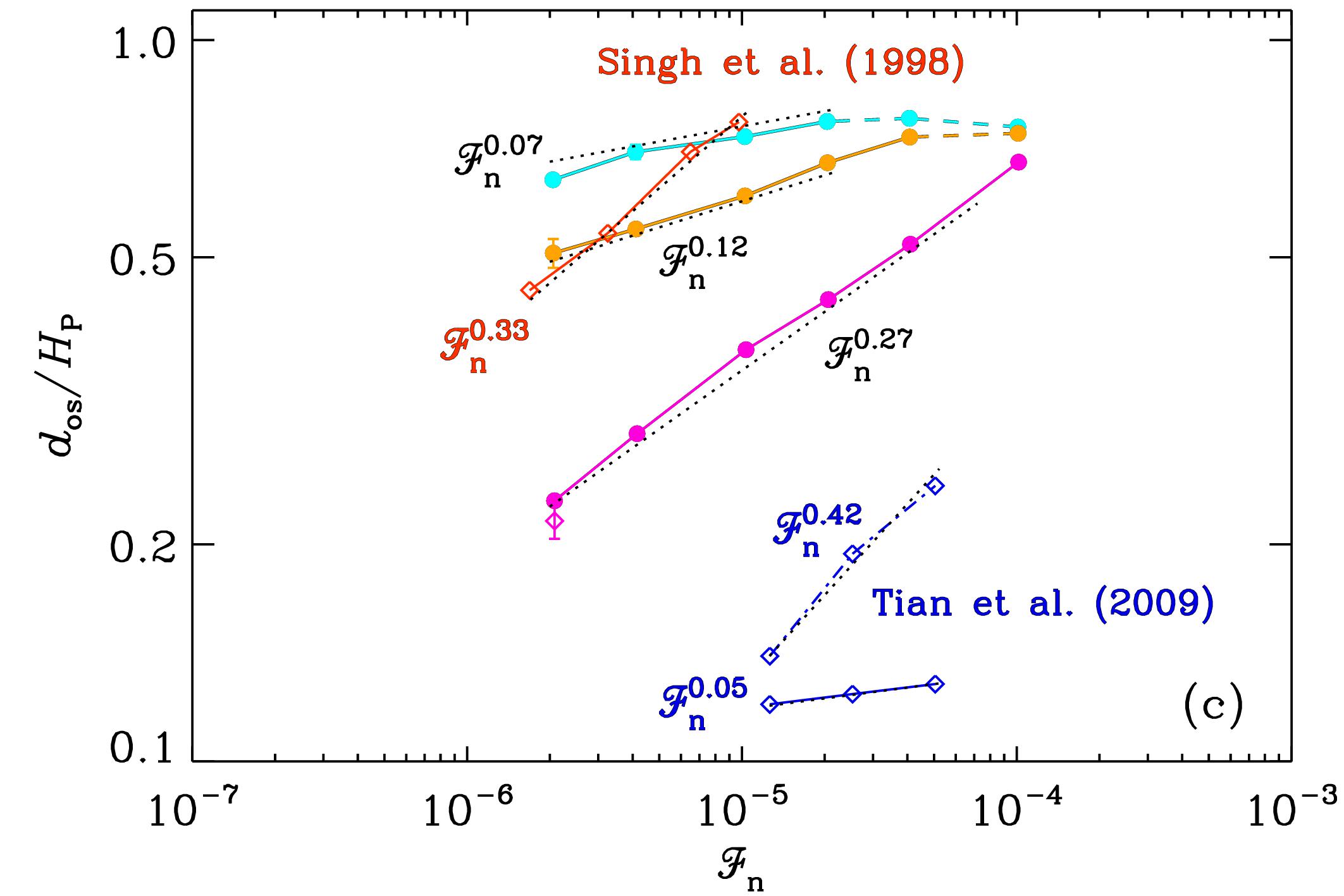}
  \caption{({\it a}) Vertical ($z$) coordinates of the bottom of the
    CZ ($z_{\rm CZ}$, solid lines) and OZ ($z_{\rm OZ}$, dashed). The
    dotted red line indicates the bottom of the domain. ({\it b})
    Overshooting depth $d_{\rm os}$ normalised by the pressure scale
    height $\Hp$ as a function of $\mathscr{F}_{\rm n}$ for Sets~K
    (black), Kh (grey), P (blue), S (red), Sh (green), DS (purple),
    DSS1 (cyan), and DSS2 (orange). The purple diamond denotes
    Run~DS5h. The dotted lines show approximate power laws from fits
    to simulation data; see \Table{tab:Pwrlaws}. ({\it c}) Comparison
    of Sets~DS, DSS1, and DSS2 with the studies of
    \cite{1998A&A...340..178S} (red) and \cite{2009MNRAS.398.1011T}
    (blue).}
\label{fig:plot_oshoot}
\end{figure}

One of the most restrictive modelling choices in the past has been the
use of a static heat conduction profile that effectively enforces the
layer structure of the simulation. This can be seen from
\Fig{fig:plot_oshoot}(a) where the vertical coordinates of the bottoms
of the convection ($z_{\rm CZ}$) and overshoot ($z_{\rm OZ}$) zones
are shown as functions of $\Fn$. The results for $z_{\rm CZ}$ from
Sets~S and Sh show that the interface between the CZ and OZ stays at
the initial position at $z=0$ for all values of $\Fn$. The runs in
Sets~DS, DSS1, and DSS2 behave similarly, although the bottom of the CZ is
shifted downward from its initial position. For Sets~K and Kh the
depth of the convection zone is generally reduced in comparison to
Sets~S and Sh. In these runs the depth of the CZ increases as $\Fn$
decreases. This is contrasted by the results from Sets~P where a
Kramers-like, but static, profile of $K$ is used (see
\Fig{fig:pKappa_sets}): here $z_{\rm CZ}$ is practically fixed in the
current range of $\Fn$.

\Fig{fig:plot_oshoot}(a) shows that the location of the base of the OZ
($z_{\rm OZ}$)  increases monotonically as a function of $\Fn$ in
all sets except for K and Kh. In Set~DSS1 (DSS2) the overshooting extends
to the lower boundary of the domain in the three (two) highest $\Fn$
runs, rendering the results of these simulations unusable in the
following analysis. The depth of the overshoot zone, $d_{\rm os}$, as
a function of $\Fn$ is shown in \Fig{fig:plot_oshoot}b. The results
for Sets~K, Kh, and P fall almost on top of each other. The data
suggests two power laws: $\Fn^{0.08}$ for $\Fn\lesssim10^{-5}$ and
$\Fn^{0.19}$ for $\Fn\gtrsim10^{-5}$; see \Table{tab:Pwrlaws}. These
results suggest that the overshooting depth is relatively insensitive
to the choice of the heat conduction scheme if the profile of $K$ at
the base of the CZ is smooth. Furthermore, $d_{\rm os}$ is
consistently greater in Set~Kh with higher Reynolds and SGS P\'eclet
numbers than in the corresponding runs in Set~K. This is
because the current simulations have relatively modest Reynolds and
P\'eclet numbers that are not in a fully turbulent regime. This aspect
is studied in more detail in \Sec{sec:diffcoef}. At first glance, the
data of Set~Sh appear to be more consistent with a single power, and in Set~S there appears to be a break around $\Fn \approx
10^{-5}$, such that the data points for lower values of $\Fn$ lie
below those of Set~Sh. However, power-law fits for both full and
partial ranges are consistent with a $\Fn^{0.12}$ scaling within the
error estimates; see \Table{tab:Pwrlaws}.

\begin{table}[t!]
  \centering
  \caption[]{Power-law exponents and standard mean errors from fitting
    $d_{\rm os} \propto \Fn^\alpha$.}
  \label{tab:Pwrlaws}
  \vspace{-0.5cm}
  $$
  \begin{array}{p{0.15\linewidth}lcc}
    \hline
    \noalign{\smallskip}
    \mbox{Set} & \Fn\ \mbox{range}  & \alpha \\ \hline
    \mbox{K}  & > 10^{-5} & 0.194\pm0.040 \\
    \mbox{K}  & < 10^{-5} & 0.082\pm0.011 \\
    \mbox{Kh} & > 10^{-5} & 0.184\pm0.037 \\
    \mbox{Kh} & < 10^{-5} & 0.078\pm0.018 \\
    \mbox{P}  & < 10^{-5} & 0.085\pm0.011 \\
    \hline
    \mbox{S}  & \mbox{full} & 0.119\pm0.030 \\
    \mbox{S}  &  < 10^{-5}  & 0.119\pm0.012 \\
    \mbox{Sh} & \mbox{full} & 0.138\pm0.038 \\
    \mbox{Sh} &  < 10^{-5}  & 0.121\pm0.015 \\
    \hline
    \mbox{DS} & \mbox{full} & 0.274\pm0.018 \\
    \mbox{DSS1} & < 2 \cdot 10^{-5} & 0.073\pm0.022 \\
    \mbox{DSS2} & < 4 \cdot 10^{-5} & 0.124\pm0.010 \\
    \hline
  \end{array}
  $$ \tablefoot{The ranges of $\Fn$ reflect the break of the power law
    around $\Fn=10^{-5}$ in Sets~K, Kh, and P. The same range is used
    also for Set~S and Sh, for which fits to the full range are also
    shown.}
\end{table}

A similar set-up as in \cite{1998A&A...340..178S} and
\cite{2009MNRAS.398.1011T} is adopted in Sets~DS, DSS1, and DSS2. The
results of Set~DS show a steep dependence of the overshooting on the
input flux ($d_{\rm os}\propto\Fn^{0.27}$), and Sets~DSS1 with
$d_{\rm os}\propto\Fn^{0.07}$ and DSS2 with $d_{\rm
  os}\propto\Fn^{0.12}$ are more in line with the other sets of
simulations. The overshooting depth for the higher resolution Run~DS5h
is statistically consistent with that of Run~DS5, although the overall
velocities in these runs are somewhat lower. This is in particular
reflected by the values of $\Rey_{\rm OZ}$ and $\Pe_{\rm OZ}^{\rm
  eff}$ (seventh and eighth columns in \Table{tab:runsDS}). The fact
that the results do not change drastically with resolution and that
there is an apparently continuous transition as $\Pra_{\rm SGS}^{(1)}$
increases through Sets~DSS1 and DSS2 to DS are indicative that the
latter set of runs is sufficiently well resolved
numerically. \Fig{fig:plot_oshoot}(c) shows a comparison of the
overshooting depths from Sets~DS, DSS1, and DSS2 with the studies of
\cite{1998A&A...340..178S} (their Table~1) and
\cite{2009MNRAS.398.1011T} (their Table~4). Both studies list the
input fluxes $F_{\rm b}$ and values of density and pressure at the
bottom of the domain (denoted here as $\rho_{\rm b}$ and $p_{\rm
  b}$). Thus the normalised flux in both cases is computed from
$\Fn=F_{\rm b}/F_{\rm n}$ , where $F_{\rm n} = \rho_{\rm b} \cs^3 =
\rho_{\rm b}^{-1/2} (\gamma p_{\rm b})^{3/2}$ with $\cs^2=\gamma
p_{\rm b}/\rho_{\rm b}$ and $\gamma=5/3$ has been assumed. The results
of \cite{1998A&A...340..178S} were obtained by a similar definition as
used in the present study based on kinetic energy flux falling below a
threshold fraction of its value at the base of the CZ. Their results
are consistent with a $d_{\rm os}\propto\Fn^{0.33}$ scaling. On the
other hand, \cite{2009MNRAS.398.1011T} obtained a very shallow
dependence with this definition ($d_{\rm os}\propto\Fn^{0.05}$) and a
much steeper one ($d_{\rm os}\propto\Fn^{0.42}$) when using a
criterion based on the enthalpy flux falling below a threshold value
of its absolute maximum in the OZ.  It is unclear why the results from
the two definitions used by \cite{2009MNRAS.398.1011T} deviate. Both
definitions were tested with the current data and the results were
found to be in fair agreement.

\begin{figure}
  \includegraphics[width=.5\textwidth]{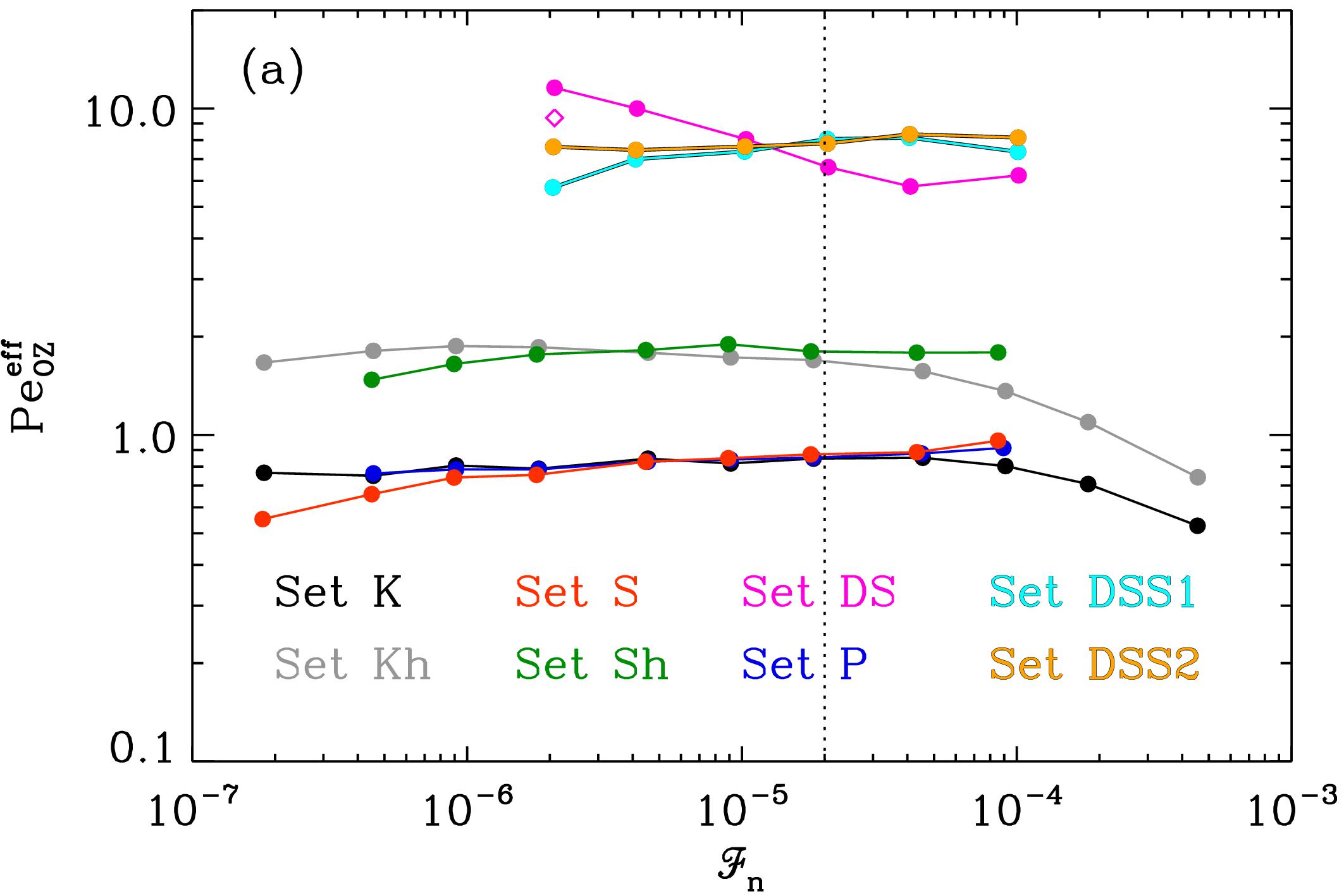}
  \includegraphics[width=.5\textwidth]{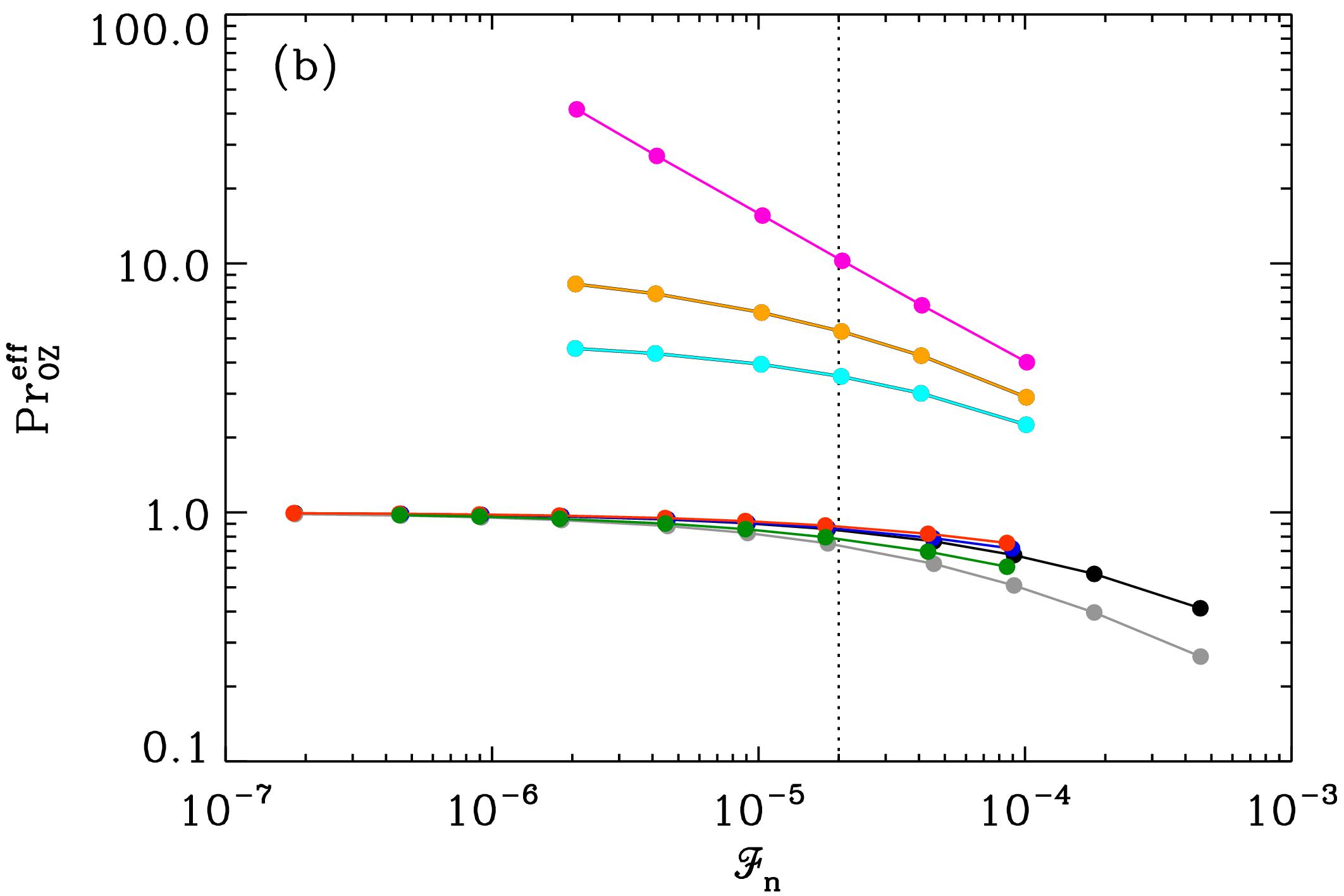}
  \caption{Effective P\'eclet {\it (a)} and Prandtl {\it (b)} numbers at the
    base of the OZ. The vertical dotted line shows the approximate
    position of the break in the power laws in the overshooting depth
    in \Fig{fig:plot_oshoot}(b).}
\label{fig:ppeclet}
\end{figure}

\begin{figure}
  \includegraphics[width=.5\textwidth]{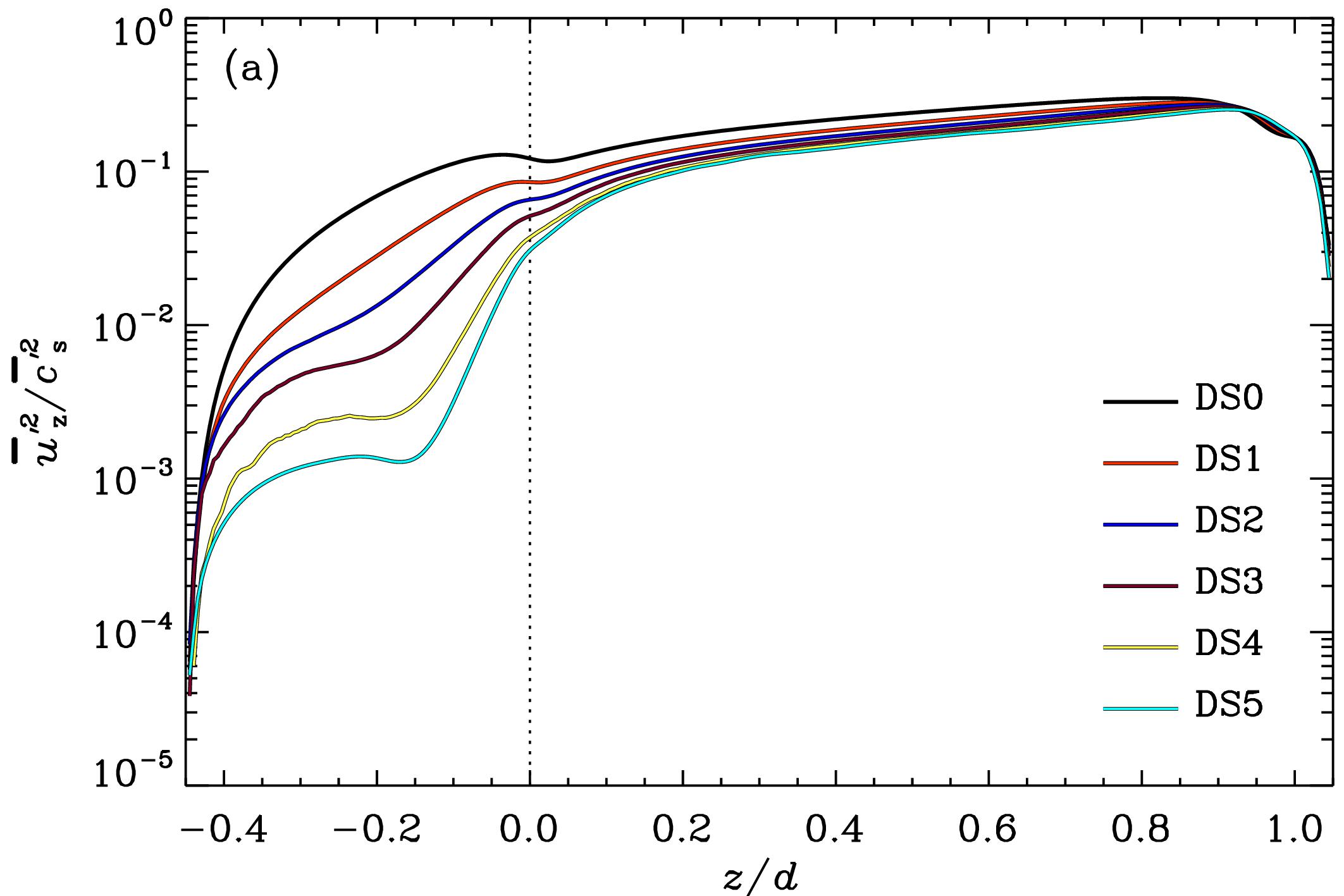}
  \includegraphics[width=.5\textwidth]{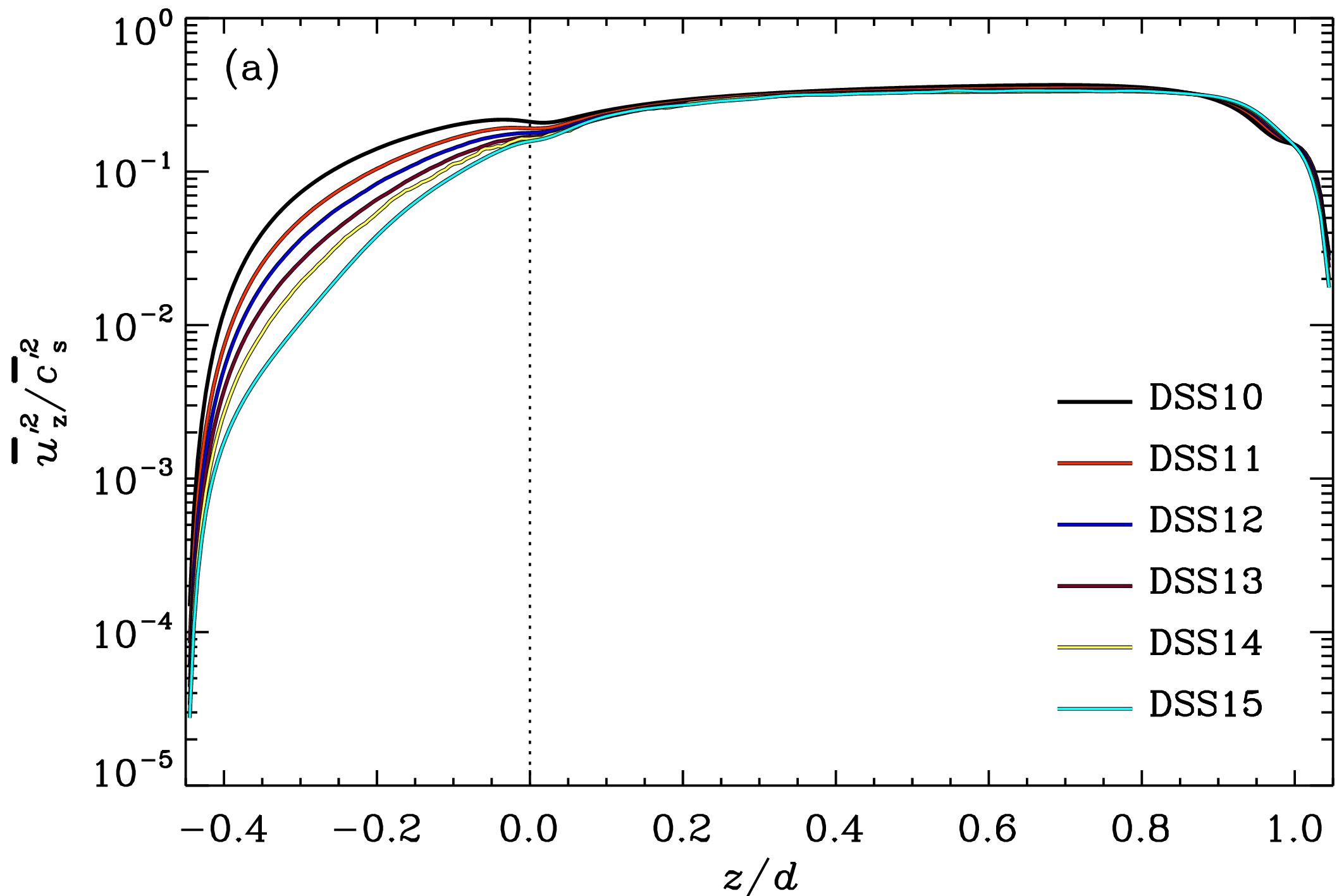}
  \caption{Ratio of the rms vertical velocity and sound speed
    fluctuations from the runs in Set~DS ({\it a}) and the runs in
    Set~DSS ({\it b}).}
\label{fig:purT}
\end{figure}

The drastic change from a steep power law in Set~DS to the
shallower power laws in Sets~DSS1 and DSS2 is related to the
difference in the diffusion of
temperature fluctuations. \Fig{fig:ppeclet}(a) shows that the
effective P\'eclet numbers in OZ in all of these sets are
comparable. However,
as is shown in \Sec{sec:diffcoef}, the overshooting depth is
insensitive to the P\'eclet number in this parameter range provided
that the effective Prandtl number is not varied at the same time. The only
difference between Sets~DS and DSS1 and DSS2 is that in
Set~DS, the SGS entropy
diffusion is omitted and thus the effective Rayleigh and Prandtl
numbers vary as a function of $\Fn$ , whereas in the latter
two, they approach constant values for low $\Fn$, see
\Figas{fig:pRayleigh}{fig:ppeclet}(b). Changing the effective Prandtl
number leads to a dramatic change in the way convection transports
energy in that the sound speed (temperature) fluctuations are enhanced
over the velocity fluctuations with increasing Prandtl number, see
\Fig{fig:purT}(a). This means that in Set~DS the temperature
fluctuations become increasingly more important in the enthalpy
transport as $\Fn$ ($\Pr$) decreases (increases). This is immediately
reflected in the overshooting depth as a smaller velocity fluctuation
is required to carry the same flux. \Fig{fig:purT}(b) shows that in
Set~DSS1 the ratio of the temperature and velocity fluctuations remains
practically constant as a function of $\Fn$. The break in the power
law in Sets~K and Kh can be understood similarly by the decreasing
effective Prandtl number as $\Fn$ increases.

To connect the current findings to earlier studies, it is necessary to
study the Prandtl number regimes explored by
\cite{1998A&A...340..178S}, \cite{2009MNRAS.398.1011T}, and
\cite{2017ApJ...843...52H}. In the two former studies a Smagorinsky
viscosity $\nu_{\rm S}$ was computed from the flow and entropy
diffusion according to $\chi_{\rm S}= {\rm Pr}_{\rm S}\nu_{\rm S}$
with a constant SGS Prandtl number of ${\rm Pr}_{\rm S} = \onethird$
was used. While this leads to the same average dependence of the
diffusion coefficients as in the present study\footnote{This is due to
  $\nu_{\rm S} = (C_k \ell)^2 \sqrt{\bm{\mathsf{S}}^2} \propto
  u'\ell$ with $u'\propto \Fn^{1/3}$, $C_k=$ const. and $\ell\propto
  \Delta x$~const., and where $\Delta x$ is the grid spacing.}, the
SGS entropy diffusion in these studies was set explicitly to zero
in radiative layers. This means that while the effective
Prandtl number in the CZ is fixed, it increases in the overshoot
layer as the flux is decreased.  Thus the sensitivity to the Prandtl
number discussed above is likely to explain the steep power laws found
by \cite{1998A&A...340..178S} and \cite{2009MNRAS.398.1011T}. On the
other hand, \cite{2017ApJ...843...52H} used a slope-limited diffusion
method where the effective diffusion coefficients are also likely to
be proportional to the gradients at small (grid) scales, that is,
$\nu_{\rm sl}\propto u'\Delta x$ and $\chi_{\rm sl}\propto s'\Delta
x$. The velocity (entropy) fluctuations scale like $\Fn^{1/3}$
($\Fn^{2/3}$) \citep[cf.][]{2018arXiv180709309K} and thus this would
lead to $\Fn^{-1/3}$ scaling for the effective Prandtl number ${\rm
  Pr}_{\rm sl}=\nu_{\rm sl}/\chi_{\rm sl}$. Whether this simplistic
picture of the effective Prandtl number with slope-limited diffusion
when applied separately for the momentum and entropy equations is
correct remains to be tested with numerical experiments. However, if
it is true, then the effective Prandtl number is increasing with
decreasing flux and is likely to contribute to the steep power law
reported by \cite{2017ApJ...843...52H}.

Furthermore, the studies of \cite{1998A&A...340..178S},
\cite{2009MNRAS.398.1011T}, and \cite{2017ApJ...843...52H} all
considered the bottom of the CZ to be fixed and given by the initial
non-convecting state. Although all of these studies used a fixed
profile for the heat conductivity, which fixes $\zcz$, it does not
necessarily stay at the same position as in the initial state;
compare, for example, the solid blue and red and green curves in
\Fig{fig:plot_oshoot}a. This is particularly clear for Sets~DS, DSS1,
and DSS2 which are similar to the set-up of
\cite{1998A&A...340..178S}. However, it is hard to assess whether such
a systematic error is present in the results of
\cite{1998A&A...340..178S}. In the simulations of
\cite{2017ApJ...843...52H} a similar issue is also possible but it
appears that this effect may be small (e.g.\ his Figure~8).

Extrapolating from the current data of the Kramers runs (Sets~K and
Kh) to solar conditions suggests that the overshooting depth for
$\Fn^\odot$ is $\mathcal{O}(0.2\Hp)$. This is in better agreement with
the constraints from helioseismology
\citep[e.g.][]{1997MNRAS.288..572B} than the estimates using the
steeper power laws of the earlier numerical studies
\citep[e.g.][]{2017ApJ...843...52H}. However, this estimate should be
considered as an upper limit because rotation and
magnetic fields, both of which reduce overshooting
\citep[e.g.][]{BCT02,2003A&A...401..433Z,KKT04}, were omitted in the
present study. The current estimate is also of the same order of
magnitude as in early analytic models of overshooting
\citep{1982A&A...113...99V,1984ApJ...282..316S,1986A&A...157..338P}
and in two-dimensional anelastic convection models with solar-like
parameters \citep{2006ApJ...653..765R}.

\subsection{Dependence on Reynolds and P\'eclet numbers}
\label{sec:diffcoef}

\begin{figure}
  \includegraphics[width=.5\textwidth]{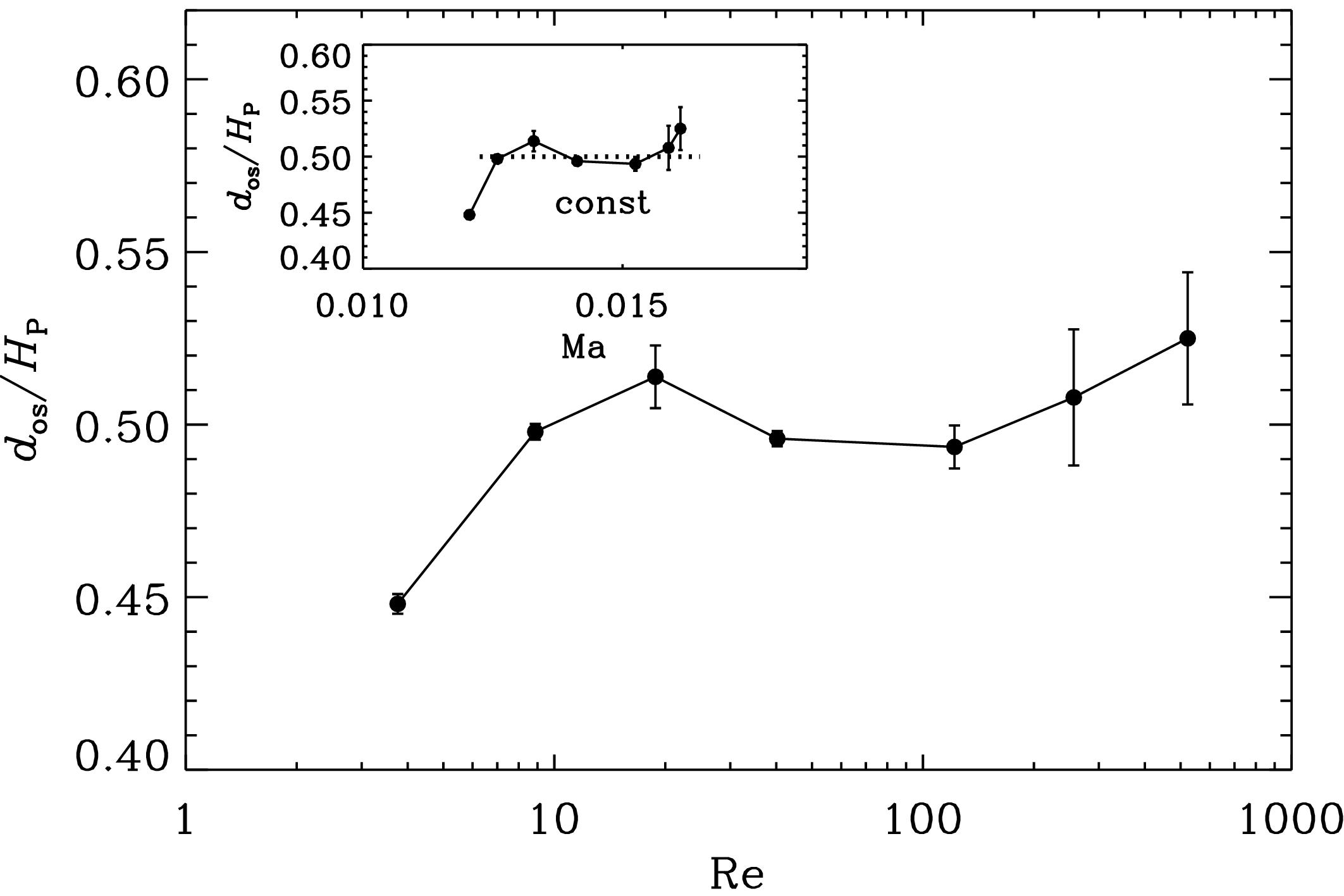}
\caption{Overshooting depth normalised by the pressure scale height at
  $z_{\rm CZ}$ as a function of $\Rey$ for Set~R. The inset shows
  $d_{\rm os}/H_{\rm P}$ as a function of $\Ma=\urms/(gd)^{1/2}$.}
\label{fig:plot_oshoot_res}
\end{figure}

In an earlier study, \cite{2017ApJ...843...52H} concluded that the
overshooting depth depends strongly on the resolution of the
simulations. The numerical models of \cite{2017ApJ...843...52H} use a
numerical diffusion scheme based on slope limiters where the effective
Reynolds and P\'eclet numbers depend on the grid spacing. Here the
explicit viscosity and entropy diffusion are varied to study this
effect. Run~K5 is taken as a reference run, and the diffusion
coefficients were varied within current computational limits in
Set~R. Run~K5 is referred to as Run~R3 in Set~R. The simulation
strategy was such that two branches of runs were performed by taking a
thermally saturated snapshot of Run~K5 as a basis. In the low-$\Rey$
branch the grid resolution was kept fixed and the diffusivities were
increased (Runs~R1 and R2). In the high-$\Rey$ branch (Runs~R4-7) the
diffusivities were decreased, and if necessary, a snapshot from a
previous simulation was re-meshed to a higher grid resolution (Runs~R5
and R6). These two branches were ran consecutively such that the
previous runs were first run to a thermally saturated state before
changing the diffusivities for the next run to avoid long transients.
The results for the normalised overshooting depth as a function of
$\Rey=\Pe_{\rm SGS}$ are shown in \Fig{fig:plot_oshoot_res}.

The current results suggest that overshooting is roughly constant as a
function of $\Rey$. We note, however, that the data points with the
highest values of $\Rey$ (Runs~R6 and R7) in \Fig{fig:plot_oshoot_res}
could not be run sufficiently long to establish that they are truly in
a statistically stationary state. Thus the values of $d_{\rm os}$ from
these runs should be considered as upper limits. In any case, these
results are at odds with those obtained by \cite{2017ApJ...843...52H},
who found a steeply declining trend as a function of $\Rey$. This,
however, is likely because \cite{2017ApJ...843...52H}
modified the heat conductivity in the radiative layer to speed up
thermal relaxation (see \Sec{sec:modK}), and possibly
  exacerbated by the varying effective Prandtl number in his models
  (\Sec{sec:flux}).

The inset of \Fig{fig:plot_oshoot_res} shows $d_{\rm os}/\Hp$ as a
function of $\Ma$ which quantifies the overall magnitude of the
convective velocity. The current data suggest that the overshooting
depth is independent of the overall velocity. This is at odds with \cite{Za91}, for instance, who derived a $\Ma^{3/2}$
dependence. However, the range of values explored here is too narrow
to draw definite conclusions.

\begin{figure}
  \includegraphics[width=.5\textwidth]{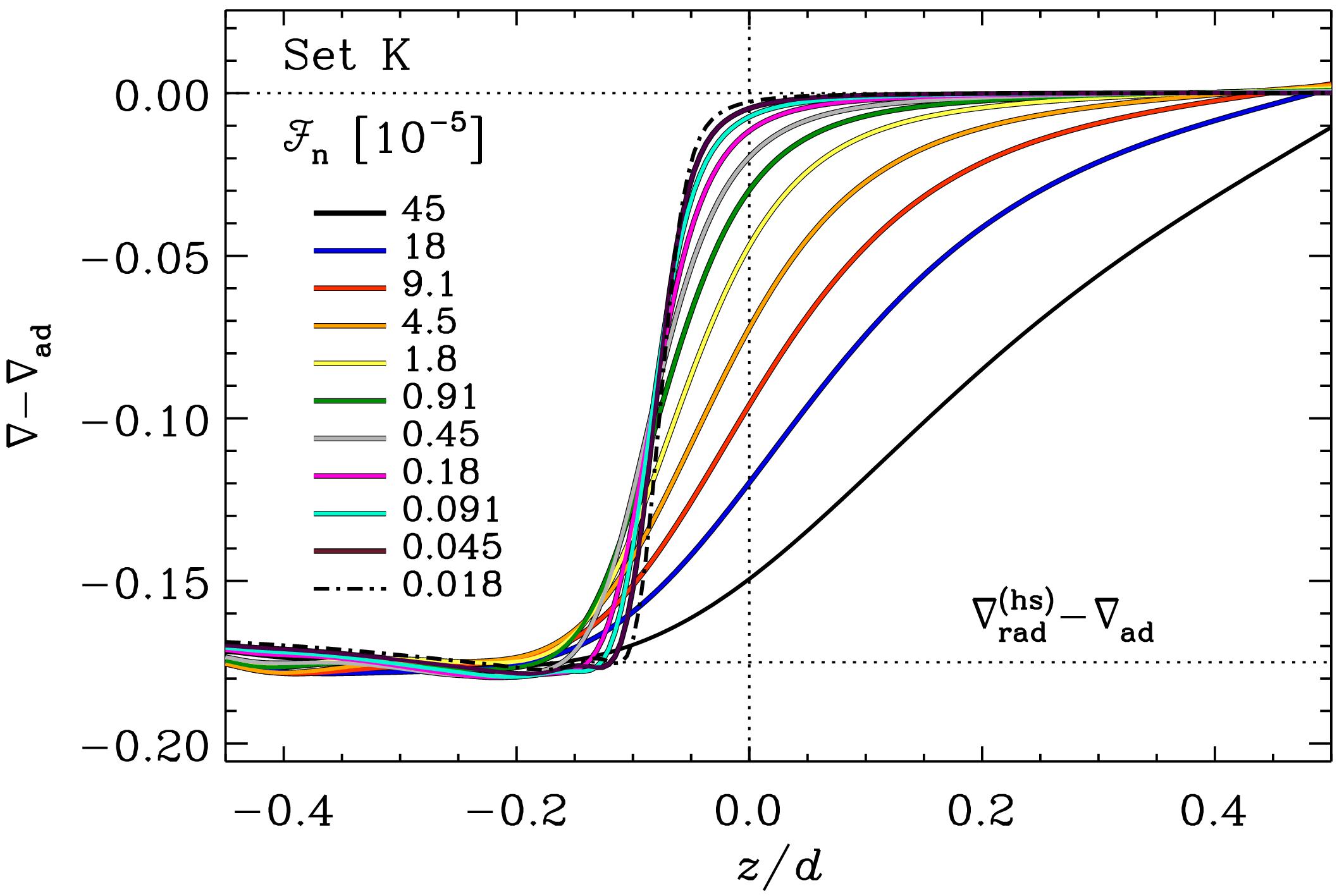}
  \caption{Superadiabatic temperature gradient at the bottom of
    the CZ in Set~K. The normalised energy flux in each
    run is indicated in the legend. The vertical dotted line indicates
    the position of the bottom of the CZ in the initial state, and
    $\nabla_{\rm rad}^{\rm (hs)}$ corresponds to the hydrostatic
    solution in the case where $\bm\nabla T = {\rm const}$.}
\label{fig:pnabla}
\end{figure}

\begin{figure}
  \includegraphics[width=.5\textwidth]{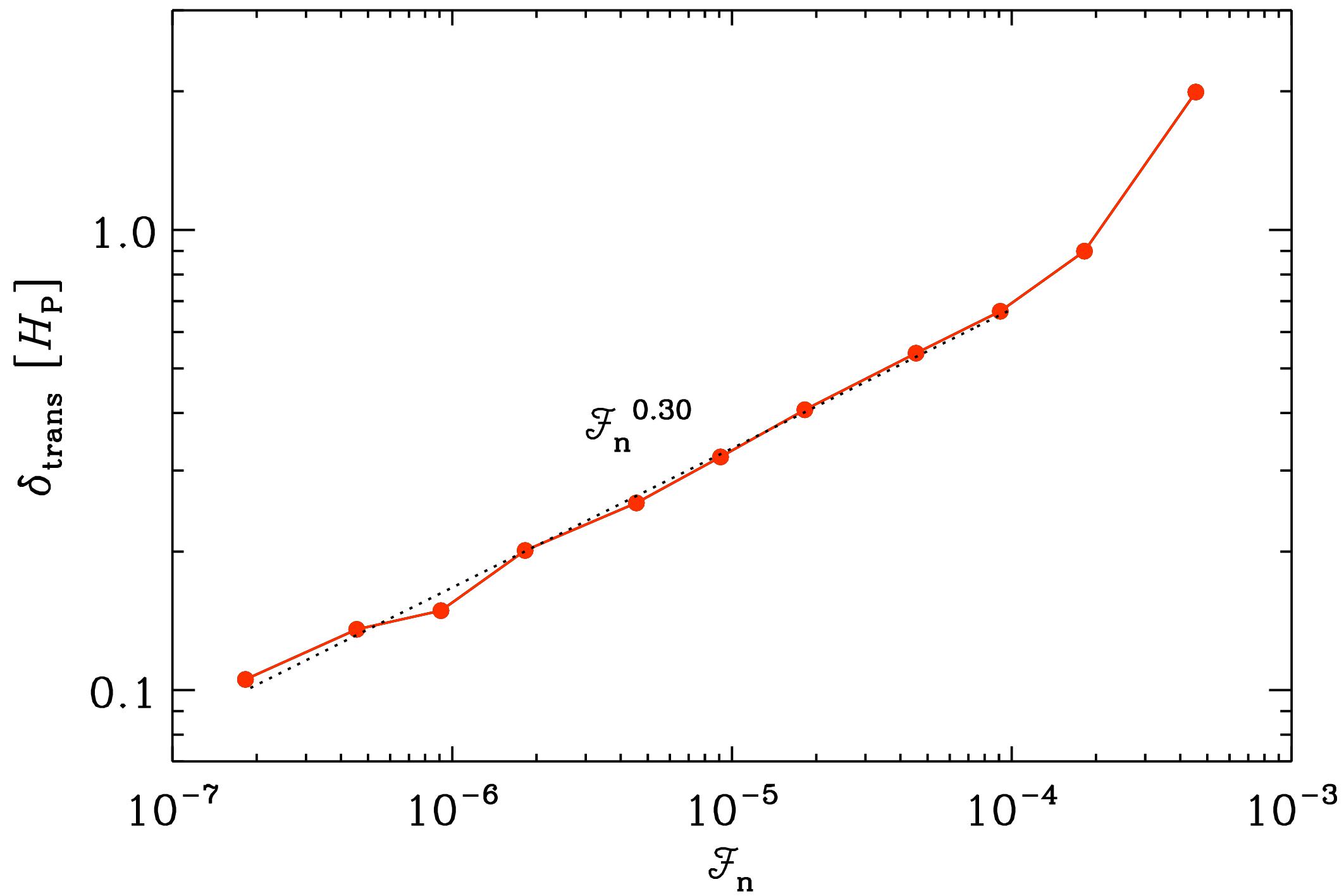}
\caption{Depth of the transition $\delta_{\rm trans}$ from nearly
  adiabatic to radiative zones as a function of input flux $\Fn$
  normalised by the pressure scale height at $z_{\rm CZ}$ from Set~K.}
\label{fig:pnabla_trans}
\end{figure}

\begin{figure}
  \includegraphics[width=.5\textwidth]{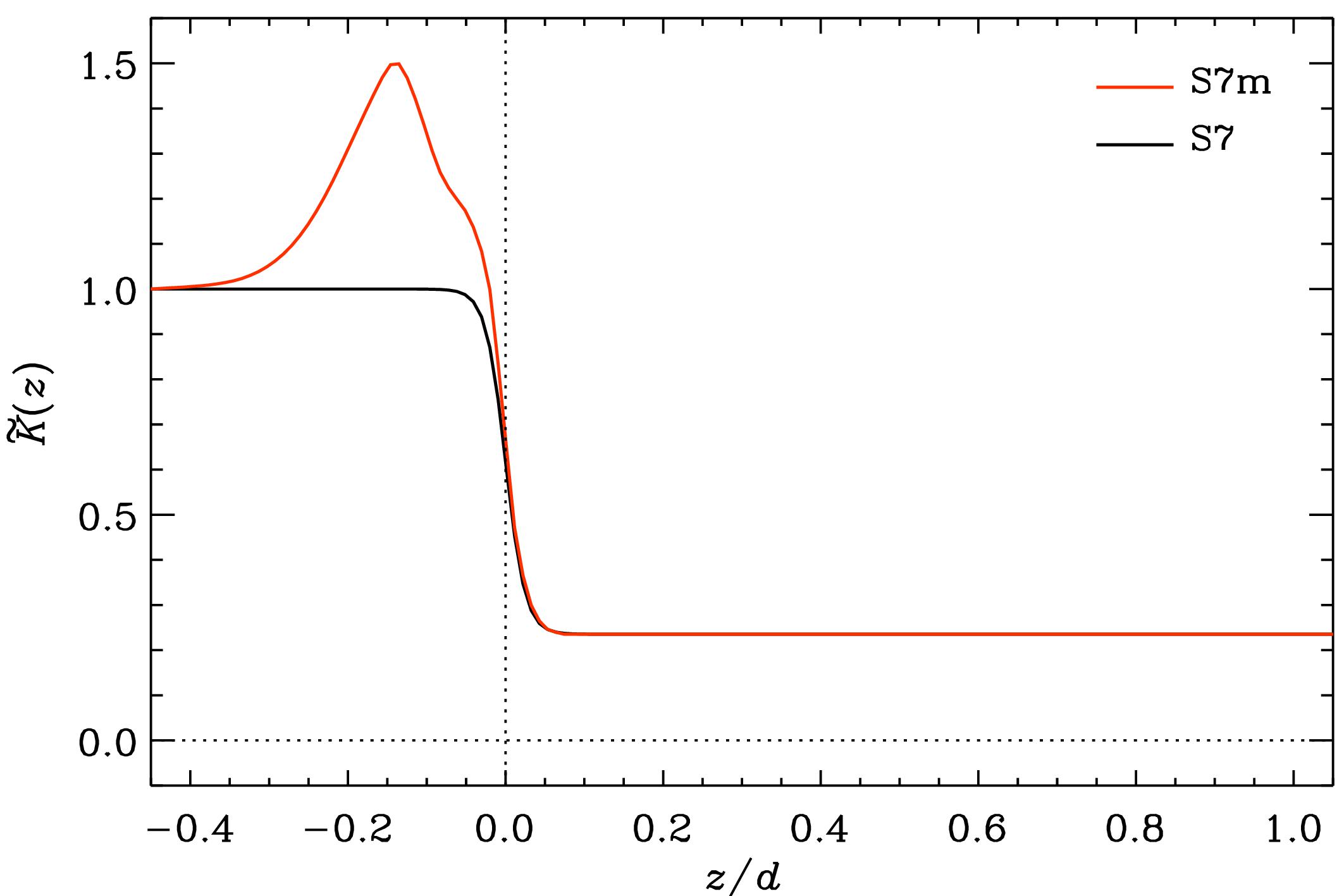}
\caption{Profiles of $K$, normalised by $K_{\rm bot}$, from Runs~S7
  (black) and S7m (red). The bottom of the initially unstable layer is
  indicated by the vertical dotted line at $z=0$.}
\label{fig:pKappa}
\end{figure}

\begin{figure*}
\sidecaption
  \includegraphics[width=.7\textwidth]{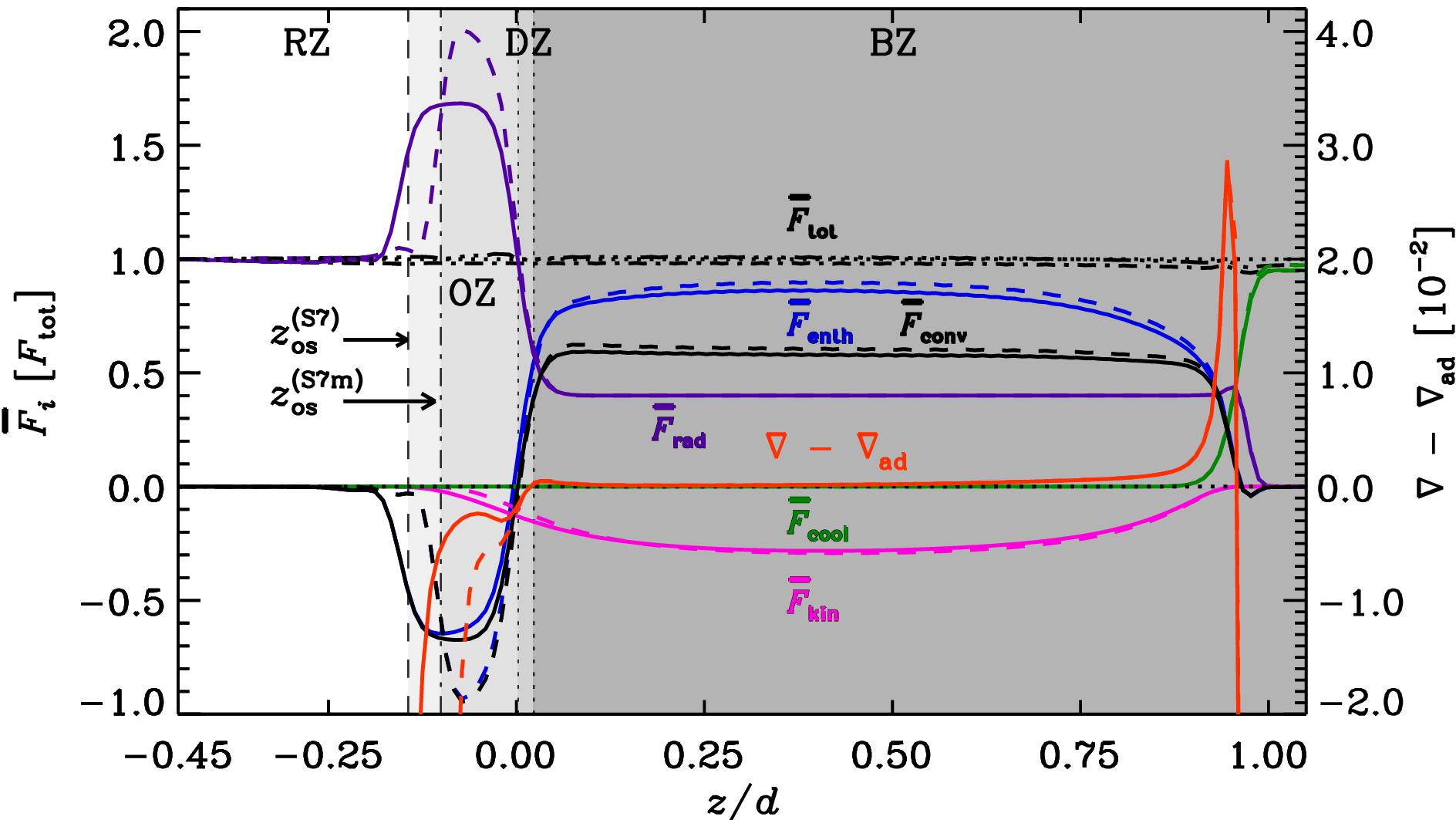}
\caption{Time-averaged total (black dash-dotted),
  convective (black), enthalpy (blue), radiative (dark
  purple), kinetic energy (light purple), and cooling (green) fluxes
  from Runs~S7 (solid) and S7m (dashed). The vertical dotted lines
  indicate the bottoms of the buoyancy (BZ) and Deardorff zones
  (DZ). The bottom of the overshoot zone (OZ) for Run~S7 (S7m) is
  denoted by the dashed (dot-dashed) vertical line.}
\label{fig:pflux_s144a7}
\end{figure*}

\subsection{Transition from the nearly adiabatic to the radiative zone}
\label{sec:trans}

The superadiabatic temperature gradient from the runs in Set~K is
shown in \Fig{fig:pnabla}. The value of $\nabla-\nabla_{\rm ad}$ in
the RZ is close to that of the hydrostatic solution with
$\nabla T=\mbox{const.}$ in a polytropic atmosphere with adiabatic
index $n=13/4$, that is, $\nabrad^{\rm (hs)}-\nabla_{\rm
  ad}=-17/85\approx-0.165$ \citep[see also][]{2019GApFD.113..149K}.
The transition from nearly adiabatic to the radiative gradient
becomes increasingly sharper as $\Fn$ decreases
\citep[e.g.][]{KKST07}. Furthermore, the temperature gradient in the
upper part of the OZ also approaches adiabatic as a
function of $\Fn$ , suggesting
penetration in the nomenclature of \cite{Za91}.

The convection zone is characterised by $\nabla\approx\nabad$ , whereas
in the radiative zone, $\nabla\approx\nabrad^{\rm (hs)}$. Thus a rough
estimate of the width of the transition between the zones and
its dependence on $\Fn$ is obtained by computing the vertical
derivative of $\nabla-\nabla_{\rm ad}$, taking its maximum, and
computing where the derivative drops below a fixed fraction of the
maximum. Here the threshold was set at half the maximum value. The
results for the computed depth of the transition $\delta_{\rm trans}$
from Set~K are shown as a function of $\Fn$ in
\Fig{fig:pnabla_trans}. The results indicate a power-law $\delta_{\rm
  trans}\propto\Fn^{0.30}$ for $\Fn\lesssim10^{-4}$. For higher values
of $\Fn$ the approximation $\nabla\approx\nabad$ is no longer accurate
in the CZ and the results deviate from the general
trend. Extrapolating from $\delta_{\rm trans}\approx0.1\Hp$ for
$\Fn=1.9\cdot 10^{-7}$, a value of $\delta_{\rm
  trans}\approx7.9\cdot10^{-3}\Hp$ is obtained for the solar value of
$\Fn\approx4\cdot10^{-11}$. This corresponds to roughly 400~km and
indicates a sharp transition between overshoot and radiative
zones. This is similar to what \cite{1984ApJ...282..316S} found based
on a non-local ML model.

\subsection{Modified heat conductivity in the radiative zone}
\label{sec:modK}

The study of \cite{2017ApJ...843...52H} reached the lowest
value of $\Fn$ in the literature thus far, with $\Fn=5\cdot10^{-7}$. However,
these results were obtained by modifying the heat conductivity in the
radiative and overshoot layers while the simulations were
running. This was done to achieve a statistically stationary state
without having to run a full thermal diffusion time. This procedure is
sometimes used in anelastic simulations in order to avoid having to
run a prohibitively long Kelvin--Helmholtz time
\citep[e.g.][]{BMT11,2017ApJ...836..192B}. While this procedure can
potentially shorten the time to saturation considerably, it can
have serious repercussions for overshooting. To demonstrate this, a
new run was branched off from Run~S7 in which the profile of
$\mean{K}(z)$ in the OZ and RZ was modified. The procedure entails
computing the energy flux from the non-relaxed run and modifying
$\mean{K}$ such that the sum of all the fluxes matches $\Ftot$ in the
RZ and OZ \citep{2017ApJ...843...52H}, that is,
\begin{eqnarray}
  \mean{K}' = \frac{\mFenth + \mFkin + \mFvisc - \Ftot}{\pd_z \mean{T}}.
\end{eqnarray}
The original and modified profiles of $\mean{K}$ for Runs~S7 and S7m
are shown in \Fig{fig:pKappa}. It is important to note that this
procedure alters a crucial system parameter of the model and that
the modification can be applied at an arbitrary time in the
non-relaxed phase of the simulation.

What happens in practice is that in the early phases of the
simulations the cooling at the surface, in combination with weak
  radiative diffusion in the CZ, drives efficient convection
that overshoots significantly into the radiative layer. This leads to
a nearly adiabatic temperature gradient and to a reduced radiative
flux in the upper part of the RZ. If $\Fn$ is low, the radiative flux
is not replenished rapidly enough and the initially vigorous
convection cannot be maintained. This leads to a long period of slow
evolution in which part of the heat coming from below is deposited in
the RZ, OZ, and DZ such that the temperature gradient gradually
steepens there to ultimately allow for the total flux to be
transmitted. In the standard scenario (Run~S7), the heat conductivity
is fixed and the temperature gradient steepens, which means that the
upper part of the RZ becomes less stiff, allowing relatively deep
overshooting, see \Fig{fig:pflux_s144a7}. The situation is exactly the
opposite in Run~S7m: the temperature gradient remains shallow and the
stratification is significantly stiffer in comparison to the case
where $\mean{K}$ was not altered.

\Figu{fig:pflux_s144a7} also shows that the modification of the heat
conductivity has serious repercussions for the overshooting depth:
$d_{\rm os}$ is reduced by roughly 30 per cent from $0.33\Hp$ in
Run~K7 to $0.23\Hp$ in Run~K7m. This result demonstrates that changing
$\mean{K}$ during the run leads to a substantial underestimation of
the overshooting depth. This is particularly relevant for the higher
resolution cases where the modification of $\mean{K}$ presumably has
to be made at an earlier stage. This might explain the strongly
decreasing overshooting depth as a function of $\Fn$ in the study of
\cite{2017ApJ...843...52H}.

\section{Conclusions}

The scaling of convective overshooting at the base of the CZ was
studied as a function of the imposed energy flux, Reynolds number, and
different heat conduction profiles and prescriptions. Using heat
conductivity based on Kramers opacity, or a similar smoothly varying,
but fixed, profile leads to a $d_{\rm os} \propto \Fn^{0.08}$
dependence for $\Fn\lesssim 10^{-5}$. Furthermore, $d_{\rm os}$ is
consistent with a constant as a function of the Reynolds and P\'eclet
number in the range $\Rey=9\ldots523$ in cases where $\PraSGSo=1$. A
somewhat steeper power, $\Fn^{0.12}$ was found in cases with a fixed
step profile for the heat conduction. These results thus indicate a
much milder dependence on the imposed energy flux than previous
studies in the literature
\citep[][]{1998A&A...340..178S,2017ApJ...843...52H}. Numerical
experiments with set-ups where the SGS diffusion was turned off led to
a steep power law ($d_{\rm os} \propto \Fn^{0.27}$) similar to the power laws
reported earlier. Otherwise identical runs with relatively weak SGS
diffusion with $\PraSGSo=5$ and 10, on the other hand, produced
shallower dependences ($d_{\rm os}\propto\Fn^{0.07}$, and
$d_{\rm os}\propto\Fn^{0.12}$, respectively). The cause for a steep
power law in the case without SGS entropy diffusion is that the
effective Prandtl number increases proportional to $\Fn$, causing the
temperature fluctuations to increase. In such cases a smaller velocity
fluctuation is needed to carry the same flux, which leads to reduced
overshooting. This is the most likely cause for the steep power laws
reported by \cite{1998A&A...340..178S} and \cite{2009MNRAS.398.1011T},
where the effective Prandtl number in the overshoot layer is likely
much larger than unity. A similar argument can tentatively be made
regarding the slope-limited diffusion used by
\cite{2017ApJ...843...52H}, but this should be tested with further
experiments.

Furthermore, the current results indicate that modifying the heat
conductivity in the layers below the convection zone
\citep[e.g.][]{2017ApJ...836..192B,2017ApJ...843...52H} leads to a
substantial underestimation of the overshooting depth. The only
way to extract reliable scaling of the overshooting depth as a
function of $\Fn$ currently is to run the simulations self-consistently to a
thermally relaxed state without modifying the system parameters such
as the heat conductivity. The present study also demonstrates the
limits of this approach in that the runs with the lowest input flux
require integration times of the order of several months even at a
relatively low resolution of $288^3$. A more promising alternative to
speeding up thermal saturation is to alter the thermodynamic
quantities instead \citep[e.g.][]{HTM86,2018PhRvF...3h3502A}. However,
even this method has its limitations, and the applicability of this
approach, for example, to rotating convection in spherical shells
remains to be demonstrated.

The current results suggest that the overshooting depth in the Sun
would be of the order of $\mathcal{O} (0.2),\Hp$ which is somewhat
higher than the canonical estimates of $(0.05\ldots0.1)\Hp$ from
helioseismology. Furthermore, the transition from overshoot to
radiative zone is expected to be abrupt and occur over a depth of
roughly $400$~km. However, the present models lack rotation and
magnetic fields, which have a significant effect on the convective
flows in the deep parts of the solar CZ. Possibly the biggest caveat
is the unrealistically large SGS Prandtl number we used in the current
study. Set-ups where these constraints are relaxed will be explored in
future publications.

\begin{acknowledgements}
  The anonymous referee is acknowledged for the constructive comments on the
  manuscript. Axel Brandenburg is acknowledged for his insightful comments on
  the manuscript. The simulations were performed using the supercomputers
  hosted by CSC -- IT Center for Science Ltd.\ in Espoo, Finland, who are
  administered by the Finnish Ministry of Education and the Gauss Center for
  Supercomputing for the Large-Scale computing project ``Cracking the
  Convective Conundrum'' in the Leibniz Supercomputing Centre's SuperMUC
  supercomputer in Garching, Germany. This work was supported in part by the
  Deutsche Forschungsgemeinschaft Heisenberg programme (grant No.\ KA
  4825/1-1) and the Academy of Finland ReSoLVE Centre of Excellence (grant
  No.\ 307411).
\end{acknowledgements}

\bibliographystyle{aa}
\bibliography{paper}

\begin{thebibliography}{73}
\expandafter\ifx\csname natexlab\endcsname\relax\def\natexlab#1{#1}\fi

\bibitem[{{Anders} {et~al.}(2018){Anders}, {Brown}, \&
  {Oishi}}]{2018PhRvF...3h3502A}
{Anders}, E.~H., {Brown}, B.~P., \& {Oishi}, J.~S. 2018, Physical Review
  Fluids, 3, 083502

\bibitem[{{Barekat} \& {Brandenburg}(2014)}]{BB14}
{Barekat}, A. \& {Brandenburg}, A. 2014, \aap, 571, A68

\bibitem[{{Basu}(1997)}]{1997MNRAS.288..572B}
{Basu}, S. 1997, \mnras, 288, 572

\bibitem[{{Bekki} {et~al.}(2017){Bekki}, {Hotta}, \&
  {Yokoyama}}]{2017ApJ...851...74B}
{Bekki}, Y., {Hotta}, H., \& {Yokoyama}, T. 2017, \apj, 851, 74

\bibitem[{{Brandenburg}(2003)}]{B03}
{Brandenburg}, A. 2003, {Computational aspects of astrophysical MHD and
  turbulence}, ed. A.~{Ferriz-Mas} \& M.~{N{\'u}{\~n}ez} (London: Taylor and
  Francis), 269

\bibitem[{{Brandenburg}(2016)}]{Br16}
{Brandenburg}, A. 2016, \apj, 832, 6

\bibitem[{{Brandenburg} {et~al.}(2005){Brandenburg}, {Chan}, {Nordlund}, \&
  {Stein}}]{BCNS05}
{Brandenburg}, A., {Chan}, K.~L., {Nordlund}, {\AA}., \& {Stein}, R.~F. 2005,
  AN, 326, 681

\bibitem[{{Brandenburg} {et~al.}(2000){Brandenburg}, {Nordlund}, \&
  {Stein}}]{2000gac..conf...85B}
{Brandenburg}, A., {Nordlund}, A., \& {Stein}, R.~F. 2000, in Geophysical and
  Astrophysical Convection, Contributions from a workshop sponsored by the
  Geophysical Turbulence Program at the National Center for Atmospheric
  Research, October, 1995. Edited by Peter A. Fox and Robert M. Kerr. Published
  by Gordon and Breach Science Publishers, The Netherlands, 2000, p. 85-105,
  ed. P.~A. {Fox} \& R.~M. {Kerr}, 85--105

\bibitem[{{Brandenburg} \& {Subramanian}(2005)}]{BS05}
{Brandenburg}, A. \& {Subramanian}, K. 2005, \physrep, 417, 1

\bibitem[{{Brown} {et~al.}(2010){Brown}, {Browning}, {Brun}, {Miesch}, \&
  {Toomre}}]{BBBMT10}
{Brown}, B.~P., {Browning}, M.~K., {Brun}, A.~S., {Miesch}, M.~S., \& {Toomre},
  J. 2010, \apj, 711, 424

\bibitem[{{Brummell} {et~al.}(2002){Brummell}, {Clune}, \& {Toomre}}]{BCT02}
{Brummell}, N.~H., {Clune}, T.~L., \& {Toomre}, J. 2002, \apj, 570, 825

\bibitem[{{Brun} \& {Browning}(2017)}]{2017LRSP...14....4B}
{Brun}, A.~S. \& {Browning}, M.~K. 2017, Liv. Rev. Sol. Phys., 14, 4

\bibitem[{{Brun} {et~al.}(2011){Brun}, {Miesch}, \& {Toomre}}]{BMT11}
{Brun}, A.~S., {Miesch}, M.~S., \& {Toomre}, J. 2011, \apj, 742, 79

\bibitem[{{Brun} {et~al.}(2017){Brun}, {Strugarek}, {Varela}, {Matt},
  {Augustson}, {Emeriau}, {DoCao}, {Brown}, \& {Toomre}}]{2017ApJ...836..192B}
{Brun}, A.~S., {Strugarek}, A., {Varela}, J., {et~al.} 2017, \apj, 836, 192

\bibitem[{{Cai}(2018)}]{2018ApJ...868...12C}
{Cai}, T. 2018, \apj, 868, 12

\bibitem[{{Canuto}(2011)}]{Ca11a}
{Canuto}, V.~M. 2011, \aap, 528, A76

\bibitem[{{Cattaneo} {et~al.}(1991){Cattaneo}, {Brummell}, {Toomre},
  {Malagoli}, \& {Hurlburt}}]{CBTMH91}
{Cattaneo}, F., {Brummell}, N.~H., {Toomre}, J., {Malagoli}, A., \& {Hurlburt},
  N.~E. 1991, \apj, 370, 282

\bibitem[{{Chan} \& {Gigas}(1992)}]{CG92}
{Chan}, K.~L. \& {Gigas}, D. 1992, \apjl, 389, L87

\bibitem[{{Deardorff}(1961)}]{1961JAtS...18..540D}
{Deardorff}, J.~W. 1961, J. Atmosph. Sci., 18, 540

\bibitem[{{Deardorff}(1966)}]{De66}
{Deardorff}, J.~W. 1966, J. Atmosph. Sci., 23, 503

\bibitem[{{Deng} \& {Xiong}(2008)}]{2008MNRAS.386.1979D}
{Deng}, L. \& {Xiong}, D.~R. 2008, \mnras, 386, 1979

\bibitem[{{Deng} {et~al.}(2006){Deng}, {Xiong}, \&
  {Chan}}]{2006ApJ...643..426D}
{Deng}, L., {Xiong}, D.~R., \& {Chan}, K.~L. 2006, \apj, 643, 426

\bibitem[{{Dobler} {et~al.}(2006){Dobler}, {Stix}, \& {Brandenburg}}]{DSB06}
{Dobler}, W., {Stix}, M., \& {Brandenburg}, A. 2006, \apj, 638, 336

\bibitem[{{Garaud} {et~al.}(2010){Garaud}, {Ogilvie}, {Miller}, \&
  {Stellmach}}]{GOMS10}
{Garaud}, P., {Ogilvie}, G.~I., {Miller}, N., \& {Stellmach}, S. 2010, \mnras,
  407, 2451

\bibitem[{{Gastine} {et~al.}(2016){Gastine}, {Wicht}, \&
  {Aubert}}]{2016JFM...808..690G}
{Gastine}, T., {Wicht}, J., \& {Aubert}, J. 2016, Journal of Fluid Mechanics,
  808, 690

\bibitem[{{Gough}(1969)}]{1969JAtS...26..448G}
{Gough}, D.~O. 1969, Journal of Atmospheric Sciences, 26, 448

\bibitem[{{Hotta}(2017)}]{2017ApJ...843...52H}
{Hotta}, H. 2017, \apj, 843, 52

\bibitem[{{Hurlburt} {et~al.}(1986){Hurlburt}, {Toomre}, \&
  {Massaguer}}]{HTM86}
{Hurlburt}, N.~E., {Toomre}, J., \& {Massaguer}, J.~M. 1986, \apj, 311, 563

\bibitem[{{Hurlburt} {et~al.}(1994){Hurlburt}, {Toomre}, {Massaguer}, \&
  {Zahn}}]{1994ApJ...421..245H}
{Hurlburt}, N.~E., {Toomre}, J., {Massaguer}, J.~M., \& {Zahn}, J.-P. 1994,
  \apj, 421, 245

\bibitem[{{K{\"a}pyl{\"a}} {et~al.}(2019{\natexlab{a}}){K{\"a}pyl{\"a}},
  {Gent}, {Olspert}, {K{\"a}pyl{\"a}}, \& {Brandenburg}}]{2018arXiv180709309K}
{K{\"a}pyl{\"a}}, P.~J., {Gent}, F.~A., {Olspert}, N., {K{\"a}pyl{\"a}}, M.~J.,
  \& {Brandenburg}, A. 2019{\natexlab{a}}, Geophysical and Astrophysical Fluid
  Dynamics, DOI:10.1080/03091929.2019.1571586

\bibitem[{{K{\"a}pyl{\"a}} {et~al.}(2007){K{\"a}pyl{\"a}}, {Korpi}, {Stix}, \&
  {Tuominen}}]{KKST07}
{K{\"a}pyl{\"a}}, P.~J., {Korpi}, M.~J., {Stix}, M., \& {Tuominen}, I. 2007, in
  IAU Symposium, Vol. 239, Convection in Astrophysics, ed. F.~{Kupka},
  I.~{Roxburgh}, \& K.~L. {Chan}, 437--442

\bibitem[{{K{\"a}pyl{\"a}} {et~al.}(2004){K{\"a}pyl{\"a}}, {Korpi}, \&
  {Tuominen}}]{KKT04}
{K{\"a}pyl{\"a}}, P.~J., {Korpi}, M.~J., \& {Tuominen}, I. 2004, \aap, 422, 793

\bibitem[{{K{\"a}pyl{\"a}} {et~al.}(2017){K{\"a}pyl{\"a}}, {Rheinhardt},
  {Brandenburg}, {Arlt}, {K{\"a}pyl{\"a}}, {Lagg}, {Olspert}, \&
  {Warnecke}}]{2017ApJ...845L..23K}
{K{\"a}pyl{\"a}}, P.~J., {Rheinhardt}, M., {Brandenburg}, A., {et~al.} 2017,
  \apjl, 845, L23

\bibitem[{{K{\"a}pyl{\"a}} {et~al.}(2019{\natexlab{b}}){K{\"a}pyl{\"a}},
  {Viviani}, {K{\"a}pyl{\"a}}, {Brandenburg}, \& {Spada}}]{2019GApFD.113..149K}
{K{\"a}pyl{\"a}}, P.~J., {Viviani}, M., {K{\"a}pyl{\"a}}, M.~J., {Brandenburg},
  A., \& {Spada}, F. 2019{\natexlab{b}}, Geophysical and Astrophysical Fluid
  Dynamics, 113, 149

\bibitem[{{Karak} {et~al.}(2015){Karak}, {K{\"a}pyl{\"a}}, {K{\"a}pyl{\"a}},
  {Brandenburg}, {Olspert}, \& {Pelt}}]{KKKBOP15}
{Karak}, B.~B., {K{\"a}pyl{\"a}}, P.~J., {K{\"a}pyl{\"a}}, M.~J., {et~al.}
  2015, \aap, 576, A26

\bibitem[{{Karak} {et~al.}(2018){Karak}, {Miesch}, \&
  {Bekki}}]{2018PhFl...30d6602K}
{Karak}, B.~B., {Miesch}, M., \& {Bekki}, Y. 2018, Physics of Fluids, 30,
  046602

\bibitem[{{Korre} {et~al.}(2017){Korre}, {Brummell}, \&
  {Garaud}}]{2017PhRvE..96c3104K}
{Korre}, L., {Brummell}, N., \& {Garaud}, P. 2017, \pre, 96, 033104

\bibitem[{{Korre} {et~al.}(2019){Korre}, {Garaud}, \&
  {Brummell}}]{2019MNRAS.484.1220K}
{Korre}, L., {Garaud}, P., \& {Brummell}, N.~H. 2019, \mnras, 484, 1220

\bibitem[{Krause \& R{\"a}dler(1980)}]{KR80}
Krause, F. \& R{\"a}dler, K.-H. 1980, {Mean-field Magnetohydrodynamics and
  Dynamo Theory} (Oxford: Pergamon Press)

\bibitem[{{Kupka}(1999)}]{1999ApJ...526L..45K}
{Kupka}, F. 1999, \apjl, 526, L45

\bibitem[{{Kupka} \& {Muthsam}(2017)}]{2017LRCA....3....1K}
{Kupka}, F. \& {Muthsam}, H.~J. 2017, Liv. Rev. Comp. Astrophys., 3, 1

\bibitem[{{Moffatt}(1978)}]{M78}
{Moffatt}, H.~K. 1978, {Magnetic Field Generation in Electrically Conducting
  Fluids} (Cambridge: Cambridge University Press)

\bibitem[{{Nelson} {et~al.}(2018){Nelson}, {Featherstone}, {Miesch}, \&
  {Toomre}}]{2018ApJ...859..117N}
{Nelson}, N.~J., {Featherstone}, N.~A., {Miesch}, M.~S., \& {Toomre}, J. 2018,
  \apj, 859, 117

\bibitem[{{Nordlund} {et~al.}(1992){Nordlund}, {Brandenburg}, {Jennings},
  {Rieutord}, {Ruokolainen}, {Stein}, \& {Tuominen}}]{NBJRRST92}
{Nordlund}, A., {Brandenburg}, A., {Jennings}, R.~L., {et~al.} 1992, \apj, 392,
  647

\bibitem[{{Ossendrijver}(2003)}]{O03}
{Ossendrijver}, M. 2003, \aapr, 11, 287

\bibitem[{{Pidatella} \& {Stix}(1986)}]{1986A&A...157..338P}
{Pidatella}, R.~M. \& {Stix}, M. 1986, \aap, 157, 338

\bibitem[{{Pratt} {et~al.}(2017){Pratt}, {Baraffe}, {Goffrey}, {Constantino},
  {Viallet}, {Popov}, {Walder}, \& {Folini}}]{2017A&A...604A.125P}
{Pratt}, J., {Baraffe}, I., {Goffrey}, T., {et~al.} 2017, \aap, 604, A125

\bibitem[{{Rempel}(2004)}]{2004ApJ...607.1046R}
{Rempel}, M. 2004, \apj, 607, 1046

\bibitem[{{Renzini}(1987)}]{1987A&A...188...49R}
{Renzini}, A. 1987, \aap, 188, 49

\bibitem[{{Rogers} {et~al.}(2006){Rogers}, {Glatzmaier}, \&
  {Jones}}]{2006ApJ...653..765R}
{Rogers}, T.~M., {Glatzmaier}, G.~A., \& {Jones}, C.~A. 2006, \apj, 653, 765

\bibitem[{{Roxburgh} \& {Simmons}(1993)}]{1993A&A...277...93R}
{Roxburgh}, L.~W. \& {Simmons}, J. 1993, \aap, 277, 93

\bibitem[{{R\"udiger}(1989)}]{R89}
{R\"udiger}, G. 1989, {Differential Rotation and Stellar Convection. Sun and
  Solar-type Stars} (Berlin: Akademie Verlag)

\bibitem[{{Saikia} {et~al.}(2000){Saikia}, {Singh}, {Chan}, {Roxburgh}, \&
  {Srivastava}}]{2000ApJ...529..402S}
{Saikia}, E., {Singh}, H.~P., {Chan}, K.~L., {Roxburgh}, I.~W., \&
  {Srivastava}, M.~P. 2000, \apj, 529, 402

\bibitem[{{Schmitt} {et~al.}(1984){Schmitt}, {Rosner}, \&
  {Bohn}}]{1984ApJ...282..316S}
{Schmitt}, J.~H.~M.~M., {Rosner}, R., \& {Bohn}, H.~U. 1984, \apj, 282, 316

\bibitem[{{Shaviv} \& {Salpeter}(1973)}]{1973ApJ...184..191S}
{Shaviv}, G. \& {Salpeter}, E.~E. 1973, \apj, 184, 191

\bibitem[{{Singh} {et~al.}(1995){Singh}, {Roxburgh}, \&
  {Chan}}]{1995A&A...295..703S}
{Singh}, H.~P., {Roxburgh}, I.~W., \& {Chan}, K.~L. 1995, \aap, 295, 703

\bibitem[{{Singh} {et~al.}(1998){Singh}, {Roxburgh}, \&
  {Chan}}]{1998A&A...340..178S}
{Singh}, H.~P., {Roxburgh}, I.~W., \& {Chan}, K.~L. 1998, \aap, 340, 178

\bibitem[{{Skaley} \& {Stix}(1991)}]{1991A&A...241..227S}
{Skaley}, D. \& {Stix}, M. 1991, \aap, 241, 227

\bibitem[{{Snellman} {et~al.}(2015){Snellman}, {K{\"a}pyl{\"a}},
  {K{\"a}pyl{\"a}}, {Rheinhardt}, \& {Dintrans}}]{2015AN....336...32S}
{Snellman}, J.~E., {K{\"a}pyl{\"a}}, P.~J., {K{\"a}pyl{\"a}}, M.~J.,
  {Rheinhardt}, M., \& {Dintrans}, B. 2015, Astron. Nachr., 336, 32

\bibitem[{{Stein} \& {Nordlund}(1989)}]{SN89}
{Stein}, R.~F. \& {Nordlund}, A. 1989, \apjl, 342, L95

\bibitem[{{Stein} \& {Nordlund}(1998)}]{1998ApJ...499..914S}
{Stein}, R.~F. \& {Nordlund}, {\AA}. 1998, \apj, 499, 914

\bibitem[{{Stix}(2002)}]{Stix02}
{Stix}, M. 2002, {The Sun: An Introduction} (Springer, Berlin)

\bibitem[{{Tian} {et~al.}(2009){Tian}, {Deng}, \& {Chan}}]{2009MNRAS.398.1011T}
{Tian}, C.-L., {Deng}, L.-C., \& {Chan}, K.-L. 2009, \mnras, 398, 1011

\bibitem[{{Tremblay} {et~al.}(2015){Tremblay}, {Ludwig}, {Freytag}, {Fontaine},
  {Steffen}, \& {Brassard}}]{2015ApJ...799..142T}
{Tremblay}, P.-E., {Ludwig}, H.-G., {Freytag}, B., {et~al.} 2015, \apj, 799,
  142

\bibitem[{{van Ballegooijen}(1982)}]{1982A&A...113...99V}
{van Ballegooijen}, A.~A. 1982, \aap, 113, 99

\bibitem[{{Vitense}(1953)}]{Vi53}
{Vitense}, E. 1953, \zap, 32, 135

\bibitem[{{Weiss} {et~al.}(2004){Weiss}, {Hillebrandt}, {Thomas}, \&
  {Ritter}}]{WHTR04}
{Weiss}, A., {Hillebrandt}, W., {Thomas}, H.-C., \& {Ritter}, H. 2004, {Cox and
  Giuli's Principles of Stellar Structure} (Cambridge, UK: Cambridge Scientific
  Publishers Ltd)

\bibitem[{{Xiong}(1985)}]{1985A&A...150..133X}
{Xiong}, D.~R. 1985, \aap, 150, 133

\bibitem[{{Yadav} {et~al.}(2016){Yadav}, {Gastine}, {Christensen}, {Duarte}, \&
  {Reiners}}]{2016GeoJI.204.1120Y}
{Yadav}, R.~K., {Gastine}, T., {Christensen}, U.~R., {Duarte}, L.~D.~V., \&
  {Reiners}, A. 2016, Geophysical Journal International, 204, 1120

\bibitem[{{Zahn}(1991)}]{Za91}
{Zahn}, J.-P. 1991, \aap, 252, 179

\bibitem[{{Zhang} {et~al.}(2012){Zhang}, {Deng}, {Xiong}, \&
  {Christensen-Dalsgaard}}]{2012ApJ...759L..14Z}
{Zhang}, C., {Deng}, L., {Xiong}, D., \& {Christensen-Dalsgaard}, J. 2012,
  \apjl, 759, L14

\bibitem[{{Zhang}(2013)}]{2013ApJS..205...18Z}
{Zhang}, Q.~S. 2013, \apjs, 205, 18

\bibitem[{{Ziegler} \& {R{\"u}diger}(2003)}]{2003A&A...401..433Z}
{Ziegler}, U. \& {R{\"u}diger}, G. 2003, \aap, 401, 433

\end{thebibliography}

\end{document}